\documentclass[aps,pre,twocolumn]{revtex4}

\usepackage{graphicx,amssymb,amsfonts,amsmath,chemarr,color,commath}

\begin{document}

\title{Diffusion vs. direct transport in the precision of morphogen readout}

\author{Sean Fancher}
\affiliation{Department of Physics and Astronomy, Purdue University, West Lafayette, IN 47907, USA}

\author{Andrew Mugler}
\email{amugler@purdue.edu}
\affiliation{Department of Physics and Astronomy, Purdue University, West Lafayette, IN 47907, USA}

\begin{abstract}
Morphogen profiles allow cells to determine their position within a developing organism, but the mechanisms behind the formation of these profiles are still not well agreed upon. Here we derive fundamental limits to the precision of morphogen concentration sensing for two canonical models: the diffusion of morphogen through extracellular space and the direct transport of morphogen from source cell to target cell, e.g. via cytonemes. We find that direct transport establishes a morphogen profile without adding extrinsic noise. Despite this advantage, we find that for sufficiently large values of population size and profile length, the diffusion mechanism is many times more precise due to a higher refresh rate of morphogen molecules. Our predictions are supported by data from a wide variety of morphogens in developing organisms.
\end{abstract}

\maketitle

Within developing organisms, morphogen profiles provide cells with information about their position relative to other cells. While it has been shown that cells can use this information to determine their position with extremely high precision \cite{dubuis2013positional,erdmann2009role,gregor2007probing,houchmandzadeh2002establishment,de2010precision}, the mechanism by which morphogen profiles are formed is still not well agreed upon. One well-known theory for such a mechanism is the synthesis-diffusion-clearance (SDC) model in which morphogen molecules are produced by localized source cells and diffuse through extracellular space before degrading or being internalized by target cells \cite{akiyama2015morphogen,gierer1972theory,lander2002morphogen,muller2013morphogen,rogers2011morphogen,wilcockson2017control}. Alternatively, a direct transport (DT) model has also been proposed where source cells create protrusions called cytonemes through which morphogen molecules can travel and be delivered directly to the target cells \cite{akiyama2015morphogen,bressloff2018bidirectional,kornberg2014cytonemes,muller2013morphogen,wilcockson2017control}. The presence of these two competing theories raises the question of whether there exists a difference in the performance capabilities between cells utilizing one or the other.

Experiments have shown that morphogen profiles display many characteristics consistent with the SDC model. The concentration of morphogen as a function of distance from the source cells has been observed to follow an exponential distribution for a variety of different morphogens \cite{driever1988bicoid,houchmandzadeh2002establishment}. The accumulation times for several morphogens in {\it Drosophila} have been measured and found to match the predictions made by the SDC model \cite{berezhkovskii2011formation}. In zebrafish, the molecular dynamics of the morphogen Fgf8 have been measured and found to be consistent with Brownian diffusion through extracellular space \cite{yu2009fgf8}. Despite these consistencies, recent experiments have lent support to the theory that morphogen molecules are transported through cytonemes rather than extracellular space. The establishment of the Hedgehog morphogen gradient in {\it Drosophila} has been seen to be highly correlated in both space and time with the formation of cytonemes \cite{bischoff2013cytonemes}, while Wnt morphogens have been found to be highly localized around cell protrusions such as cytonemes \cite{huang2015myoblast,stanganello2016role}. Theoretical studies of both the SDC and DT models have examined these measurable effects \cite{berezhkovskii2011formation,bressloff2018bidirectional,shvartsman2012mathematical,teimouri2015new,teimouri2016mechanisms}, but direct comparisons between the two models have thus far been poorly explored. In particular, it remains unknown whether one model allows for a cell to sense its local morphogen concentration more precisely than the other given biological parameters such as the number of cells or characteristic length scale of the morphogen profile.

Here we derive fundamental limits to the precision of morphogen concentration sensing for both the SDC and DT models. We focus on the stochastic noise caused by production, transport, and degradation of the morphogen molecules in the steady state regime. Several past studies have focused on the dynamics of morphogen profiles and their accumulation times \cite{berezhkovskii2011formation,bressloff2018bidirectional,shvartsman2012mathematical,teimouri2015new,teimouri2016mechanisms}. Here, we model morphogen profiles in the steady state regime as many of the experimental measurements to which we will later compare our results were taken during stages when the steady state approximation is valid \cite{grimm2010modelling,gregor2007stability,kicheva2007kinetics,yu2009fgf8,kanodia2009dynamics}.

Intuitively one might expect the DT model to have less noise due to the fact that molecules are directly deposited and cannot be lost to the surrounding environment. Surprisingly, we show below that for sufficiently large lengths of the morphogen profile, the SDC model produces less noise due to it being able achieve a higher effective independent measurement count. This result holds for one-, two-, and three-dimensional geometries. We compare our results with measurements of the above morphogens and find quantitative consistency with our predictions, suggesting that readout precision plays an important role in determining the mechanisms of morphogen profile establishment.

\begin{figure}
\begin{center}
\includegraphics[width=\columnwidth]{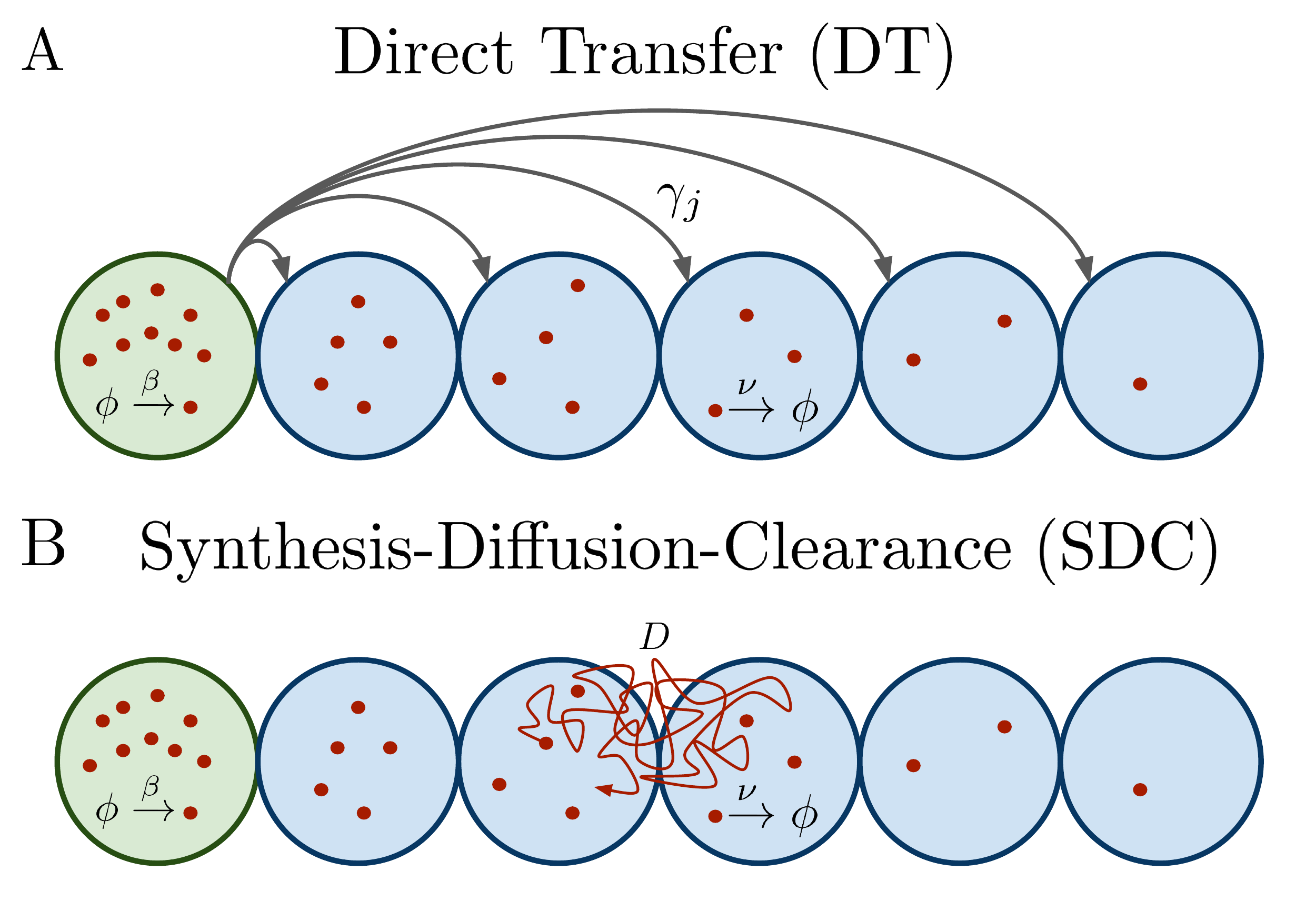}
\end{center}
\caption{Source cell (green) producing morphogen which is delivered to the target cells (blue) via direct transport (A) or diffusion (B).}
\label{fig1}
\end{figure}

We begin by considering a simple version of the DT model in which there is a single source cell that produces morphogen at rate $\beta$ (Fig.~\ref{fig1}A). Morphogen molecules are then transported to the $j$th target cell through the cytonemes, a dynamic process which we assume can be well approximated in steady state by a constant rate $\gamma_{j}$. The morphogen then degrades within the target cells at rate $\nu$. Letting $N$ be the total number of target cells, the dynamics of the number of morphogen molecules in the source cell $m_{0}\left(t\right)$ and the number in the $j$th target cell $m_{j}\left(t\right)$ are
\begin{align}
\label{m0dyn}
\frac{\partial m_{0}}{\partial t} &= \beta+\eta_{\beta}-\sum_{j=1}^{N}\left(\gamma_{j}m_{0}+\eta_{\gamma,j}\right) \\
\label{mjdyn}
\frac{\partial m_{j}}{\partial t} &= \gamma_{j}m_{0}+\eta_{\gamma,j}-\nu m_{j}-\eta_{\nu,j}.
\end{align}
The first two terms on the righthand side of Eq.~\ref{m0dyn} represent the production of morphogen and stochastic noise inherent within that process. The remaining two terms represent the transport of morphogen to the target cells and its noise. The same two terms also appear as the first terms on the righthand side of Eq.~\ref{mjdyn}, while the last two terms represent the morphogen degradation and its noise. The noise terms obey $\left\langle\eta_{\beta}\left(t'\right)\eta_{\beta}\left(t\right)\right\rangle = \beta\delta\left(t-t'\right)$, $\left\langle\eta_{\gamma,k}\left(t'\right)\eta_{\gamma,j}\left(t\right)\right\rangle = \gamma_{j}\bar{m}_{0}\delta_{jk}\delta\left(t-t'\right)$, and $\left\langle\eta_{\nu,k}\left(t'\right)\eta_{\nu,j}\left(t\right)\right\rangle = \nu\bar{m}_{j}\delta_{jk}\delta\left(t-t'\right)$ respectively \cite{gillespie2000chemical}. Here $\bar{m}_{0}=\beta/\Gamma$ and $\bar{m}_{j}=\beta\gamma_{j}/\Gamma\nu$, with $\Gamma=\sum_{j=1}^{N}\gamma_{j}$, are the steady state mean number of morphogen molecules in each cell.

From here, we assume that each cell integrates its morphogen molecule count over a time $T$ \cite{berg1977physics} such that $\left(\delta m_{j}\right)^{2}$ is the variance in the time average $T^{-1}\int_{0}^{T}dt\,m_{j}\left(t\right)$. Since Eqs.~\ref{m0dyn} and \ref{mjdyn} are linear with Gaussian white noise, solving for $\left(\delta m_{j}\right)^{2}$ is straightforward: we Fourier transform Eqs.~\ref{m0dyn} and \ref{mjdyn} in space and time, calculate the power spectrum of $m_{j}$, and approximate $\left(\delta m_{j}\right)^{2}$ by its low-frequency limit \cite{supp}. This low-frequency approximation assumes that $T\gg\{\tau_{1},\tau_{2}\}$, where $\tau_{1}=\Gamma^{-1}$ and $\tau_{2}=\nu^{-1}$ are the characteristic timescales of the morphogen molecules set by their transport and degradation rates respectively. For the $j$th target cell, this procedure yields a relative error of
\begin{equation}
\left(\frac{\delta m_{j}}{\bar{m}_{j}}\right)^{2} = \frac{2}{\nu T}\frac{1}{\bar{m}_{j}}.
\label{mjNSR}
\end{equation}
The two factors in Eq.~\ref{mjNSR} have an intuitive interpretation: the relative error is seen to decrease with both the mean number of molecules $\bar{m}_{j}$ and the number of independent measurements $T/\tau_{2}$ made in time $T$, where $\tau_{2}$ is the autocorrelation time. One important aspect of Eq.~\ref{mjNSR} is that it is identical to the case in which there is no extrinsic noise inherited from $m_{0}$. This is because the cytoneme transport process is modelled as a simple transfer action. Since no molecules are lost, any noise $m_{j}$ inherits from $m_{0}$ is cancelled by the negative cross correlations between the two \cite{supp}. The net result is that the precision of morphogen readout in any target cell is equivalent to that of a simple Poisson process, with no extra noise inherited from production and transport.

We now consider the SDC model (Fig.~\ref{fig1}B). As most morphogen systems are modelled in one-dimensional space within the SDC model \cite{berezhkovskii2011formation,shvartsman2012mathematical,teimouri2016mechanisms}, we will begin with this assumption, although we also consider other geometries later on. Consider a system with a single source cell located at the origin which produces morphogen molecules at a rate $\beta$. These molecules then diffuse freely through space with diffusion coefficient $D$ to produce a density field of morphogen, $\rho\left(x,t\right)$. The morphogen also spontaneously degrades at any point in space with rate $\nu$. The dynamics of $\rho$ are
\begin{equation}
\frac{\partial \rho}{\partial t} = D\nabla^{2}\rho+\eta_{D}-\nu\rho-\eta_{\nu}+\left(\beta+\eta_{\beta}\right)\delta\left(x\right).
\label{rhodyn}
\end{equation}
The first two terms on the righthand side of Eq.~\ref{rhodyn} represent the morphogen diffusion and noise inherent within the diffusion process. The third and fourth terms represent the degradation and its noise, and the last terms represent the production and its noise isolated to the location of the source cell. The noise terms obey $\left\langle\eta_{D}\left(x',t'\right)\eta_{D}\left(x,t\right)\right\rangle = 2D\delta\left(t-t'\right)\vec{\nabla}_{x}\cdot\vec{\nabla}_{x'}\bar{\rho}\left(x\right)\delta\left(x-x'\right)$, $\left\langle\eta_{\nu}\left(x',t'\right)\eta_{\nu}\left(x,t\right)\right\rangle = \nu\bar{\rho}\left(x\right)\delta\left(t-t'\right)\delta\left(x-x'\right)$, and $\left\langle\eta_{\beta}\left(t'\right)\eta_{\beta}\left(t\right)\right\rangle = \beta\delta\left(t-t'\right)$ respectively \cite{gardiner2004Hosm, fancher2017fundamental, varennes2017emergent}. Here $\bar{\rho}\left(x\right)$ is the steady state mean concentration of morphogen molecules at position $x$, which for a one-dimensional system with a single source takes the form $\bar{\rho}\left(x\right)=\beta\lambda\text{exp}\left(-\abs{x}/\lambda\right)/2D$ with $\lambda=\sqrt{D/\nu}$.

We then imagine a target cell located at $x$ that is perfectly permeable to the morphogen and counts the number $m\left(x,t\right) = \int_{V}dy\,\rho\left(x+y,t\right)$ of morphogen molecules within its volume $V$ as a measure of its local concentration. We use this simpler prescription over explicitly accounting for more complicated structures such as surface receptor binding because it has been shown that it ultimately yields similar concentration sensing results up to a factor of order unity \cite{berg1977physics}. Assuming $\abs{x}\ge 2a$ with $a$ being the cell radius enforces the condition that there is no overlap between the source and target cell and produces a mean value of $\bar{m}\left(x\right)=\beta\sinh\left(a/\lambda\right)\text{exp}\left(-\abs{x}/\lambda\right)/\nu$. Once again defining $\left(\delta m\left(x\right)\right)^{2}$ as the variance of the time average of $m\left(x,t\right)$ requires $T\gg\tau_{3}$ with $\tau_{3}=\left(\nu+D/a^{2}\right)^{-1}$ to make the low-frequency approximation applicable. With this, we calculate $\left(\delta m\left(x\right)\right)^{2}$ to be \cite{supp}
\begin{equation}
\left(\frac{\delta m\left(x\right)}{\bar{m}\left(x\right)}\right)^{2} = \frac{2}{\nu T}\frac{1}{\bar{m}\left(x\right)}\left(1-e^{-\frac{1}{\hat{\lambda}}}\frac{\frac{2}{\hat{\lambda}}+\sinh\left(\frac{2}{\hat{\lambda}}\right)}{4\sinh\left(\frac{1}{\hat{\lambda}}\right)}\right),
\label{mxNSR}
\end{equation}
where $\hat{\lambda}=\lambda/a$. Unlike in Eq.~\ref{mjNSR}, Eq.~\ref{mxNSR} has no negative cross correlation terms to cancel the extrinsic noise. Instead, the noise is comprised of a combination of three positive terms coming from the production, diffusion, and degradation noise terms seen in Eq.~\ref{rhodyn} \cite{supp}. These terms combine to produce the compact form in Eq.~\ref{mxNSR}. Eq.~\ref{mxNSR} has the same prefactor seen in Eq.~\ref{mjNSR} but has an additional factor that is necessarily less than 1. This additional factor is due to the fact that morphogen can diffuse away from the cell as well as degrade, which in turn means $\nu T$ is an underestimate of the number of effectively independent measurements. The additional factor compensates for this underestimation.

\begin{figure}
\begin{center}
\includegraphics[width=\columnwidth]{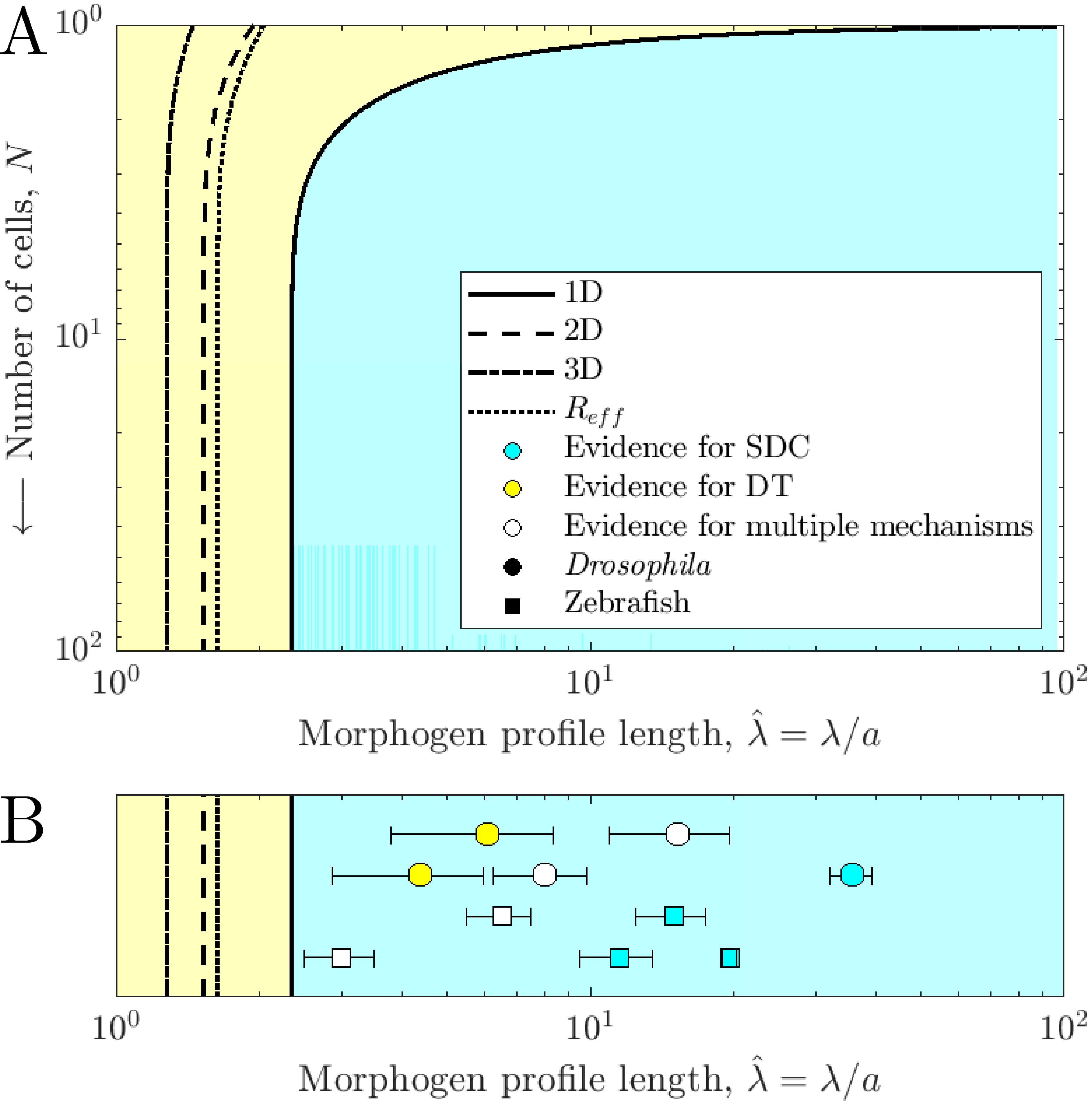}
\end{center}
\caption{(A) Plot of the boundary separating the regime in which the DT model produces lesser noise (yellow) from that in which the SDC model does (cyan) for a 1D system. Additional curves show the same boundary for 2D and 3D systems as well as for the ratio of effective number of independent measurements $R_{\text{eff}}$. (B) $\hat{\lambda}$ values for several morphogens found in {\it Drosophila} (circles) and zebrafish (squares) with boundaries from (A) assuming large $N$. Yellow markers denote morphogens which have been argued to be transported via cytonemes while blue markers denote those which have been argued to utilize diffusion. White markers denote morphogens for which there is evidence for multiple competing mechanisms. Circle markers denote (from left to right) Wg, Hh, Dorsal, Dpp, and Bicoid. Square markers denote (from left to right) Cyclops, Squint, Lefty1, Lefty2, and Fgf8. Measurements of $\hat{\lambda}$ values are from \cite{kicheva2012investigating,kicheva2007kinetics,wartlick2011dynamics,gregor2007stability,liberman2009quantitative,yu2009fgf8,muller2012differential}; see \cite{supp} for further discussion on experimental data.}
\label{fig2}
\end{figure}

By comparing Eqs.~\ref{mjNSR} and \ref{mxNSR}, it is possible to determine which model achieves a lower variance and, in turn, higher precision. We take an example system in which there are $N$ target cells in a line such that the position of the $j$th target cell is $x_{j}=2aj$. This creates a line of $N+1$ adjacent cells with the first one being the source of the morphogen, as seen in Fig.~\ref{fig1}. By setting $\gamma_{j}=\gamma_{0}\text{exp}\left(-2j/\hat{\lambda}\right)$ for any given basal transport rate $\gamma_{0}$, the mean $m$ values of both the SDC and DT models can be made to scale identically. These means can be made precisely equal to each other by appropriate choices of $\beta$ and $\nu$ in each system. We assume the $\beta$ values in both models are equivalent, as production from the source cell is a mechanism common to both models. We then fix the $\nu$ values to achieve equality in the means. With these choices made, we define $R$ to be the ratio of Eq.~\ref{mxNSR} to Eq.~\ref{mjNSR}, which yields
\begin{equation}
R = \frac{1}{e^{-\frac{N+1}{\hat{\lambda}}}\sinh\left(\frac{N}{\hat{\lambda}}\right)}\left(1-e^{-\frac{1}{\hat{\lambda}}}\frac{\frac{2}{\hat{\lambda}}+\sinh\left(\frac{2}{\hat{\lambda}}\right)}{4\sinh\left(\frac{1}{\hat{\lambda}}\right)}\right).
\label{R}
\end{equation}
When $R$ is less (greater) than 1, the SDC (DT) model achieves lesser error and is thus the more precise method of morphogen sensing. Fig.~\ref{fig2}A shows the boundary in ($N$,$\hat{\lambda}$) space that separates these two regimes (solid curve). We also calculate this boundary for a three-dimensional system assuming a plane of source cells (dot-dashed) and numerically solve for this boundary in a two-dimensional system assuming a line of source cells (dashed), both of which also produce exponentially decaying mean morphogen profiles \cite{supp}.

As can be seen in Fig.~\ref{fig2}A, for sufficiently large population size $N$ and any value of $\hat{\lambda}$ larger than $\sim$$2.35$, the SDC model is more precise. This is surprising because unlike the SDC model, the DT model carries no extrinsic noise. Additionally, in the SDC model morphogen molecules diffuse into all of space and may be ``lost" in the sense that they could require effectively infinite amounts of time to reach the target cell or, in the case of 3D space, may simply never reach it at all. These effects might suggest that the DT model should be more precise. However, our results contradict this intuition for sufficiently large values of $N$ and $\hat{\lambda}$.

As suggested by Eq.~\ref{mxNSR}, we can understand this result by considering how quickly the morphogen molecules in a target cell are refreshed. In the case of the DT model, morphogen is cleared with rate $\nu$, meaning that in a time $T$ the cell can make $\sim\nu_{\text{DT}} T$ independent measurements of its morphogen concentration. In the SDC model, morphogen can be cleared via degradation at rate $\nu$ or by simply diffusing away with an effective rate $D/a^{2}$, meaning that a cell can make $\sim\left(\nu_{\text{SDC}}+D/a^{2}\right)T$ independent measurements. We define $R_{\text{eff}}$ to be the ratio of the number of measurements made in the DT model to that made in the SDC model. Using the same parameter relations as in Eq.~\ref{R}, including the ratio of $\nu$ values between the two models, yields
\begin{equation}
R_{\text{eff}} = \frac{\nu_{\text{DT}} T}{\left(\nu_{\text{SDC}}+D/a^{2}\right)T}
= \frac{1}{e^{-\frac{N+1}{\hat{\lambda}}}\sinh\left(\frac{N}{\hat{\lambda}}\right)\left(1+\hat{\lambda}^{2}\right)}.
\label{I}
\end{equation}
As with $R$, when $R_{\text{eff}}$ is less (greater) than 1, the SDC (DT) model achieves a larger number of independent measurements and would thus be more likely to achieve a higher precision. The boundary separating these regimes is plotted in Fig.~\ref{fig2}A (dotted) and is seen to be similar to the boundary separating the $R$ regimes. As can also be seen in Fig.~\ref{fig2}A, each boundary converges to its own critical value of $\hat{\lambda}$, which we denote as $\hat{\lambda}^{c}$, as $N$ increases. Since this convergence happens quickly, we focus on the $N\to\infty$ case and ignore systems with a size on the order of only a few cells. This allows us to make the prediction that any exponential morphogen profile in which $\hat{\lambda}>\hat{\lambda}^{c}$ should use the SDC model over the DT model as it provides more precise concentration sensing, and vice versa for $\hat{\lambda}<\hat{\lambda}^{c}$.

We now test this prediction for various morphogens. In {\it Drosophila}, the morphogen Wingless has been shown to be localized near cell protrusions such as cytonemes \cite{huang2015myoblast,stanganello2016role}, while the Hedgehog gradient has been observed to correlate very highly in both space and time with the formation of cytonemes \cite{bischoff2013cytonemes}. These results suggest that these two morphogen profiles could be formed via a method akin to the DT model presented here, and as seen in Fig.~\ref{fig2}B, they both have relatively low values of $\hat{\lambda}$ \cite{kicheva2012investigating,kicheva2007kinetics,wartlick2011dynamics} and exist near the regime in which we predict the DT model is more precise than the SDC model (yellow circles). Conversely, Bicoid has been used as a model example of SDC for decades and shown to provide cells with high levels of precision in their positional information \cite{driever1988bicoid,gregor2007probing,houchmandzadeh2002establishment}. This is consistent with the fact that the Bicoid gradient has also been measured to have a very large $\hat{\lambda}$ value \cite{kicheva2012investigating,gregor2007stability}, putting it far into the regime where we predict the SDC model achieves a higher precision than the DT model (blue circle). The gradient sizes of Dorsal and Dpp have also been measured \cite{kicheva2012investigating,kicheva2007kinetics,liberman2009quantitative} and are less deeply within in the regime where we predict the SDC model is more precise (white circles), although for Dpp there is evidence for a variety of different gradient formation mechanisms \cite{akiyama2015morphogen,muller2013morphogen,wilcockson2017control}.

In zebrafish, the morphogen Fgf8 has been studied at the single molecule level and found to have molecular dynamics closely matching those expected of Brownian movement and SDC model \cite{yu2009fgf8}. The gradient size of Fgf8 has also been measured \cite{kicheva2012investigating,yu2009fgf8} and found to be deeply in our predicted SDC regime, as seen in Fig.~\ref{fig2}B (rightmost blue square). Similarly, Cyclops, Squint, Lefty1, and Lefty2, all of which are involved in the Nodal/Lefty system, have been shown to spread diffusively and be able to affect cells distant from their source \cite{muller2013morphogen,rogers2018nodal} (other squares). This would support a gradient formation mechanism similar to the SDC model although measurements of the size of Cyclops and Squint gradients \cite{kicheva2012investigating,muller2012differential} put them closer to the regime where we predict that the DT model should be more precise (white squares). This is consistent with the fact that Cyclops and Squint have been argued to be tightly regulated via a Gierer-Meinhardt type system, thus diminishing their gradient sizes to values much lower than what they would be without this regulation \cite{gierer1972theory,rogers2018nodal}.

Taken together, the results in Fig.~\ref{fig2}B support our prediction that profiles with short length scales should form via DT, whereas profiles with long length scales should form via SDC. Morphogens with experimental evidence for DT fall within a factor of two or three from our predicted DT regime, which is reasonable given the simplicity of our models, while morphogens with experimental evidence for SDC fall squarely within our predicted SDC regime.

We have shown that in the steady-state regime, the SDC and DT models of morphogen profile formation yield different scalings of readout precision with the length of the profile and population size. As a result, there exist regimes in this parameter space in which either mechanism is more precise. While the DT model inherits no extrinsic noise from the morphogen production in the source cell and molecules are never lost to diffusion, the ability of molecules to diffuse into and away from a target cell in the SDC model allows the cell to make a greater number of effectively independent measurements in the same time frame. By examining how these phenomena affect the cells' sensory precision, we predicted that morphogen profiles with lengths shorter than a critical value should utilize cytonemes or some other form of direct transport mechanism, whereas morphogens with longer profiles should rely on extracellular diffusion, a prediction that is largely in quantitative agreement with measurements on known morphogens. It will be interesting to observe whether this trend is further strengthened as more experimental evidence is obtained for different morphogens as well as to expand the theory of morphogen gradient sensing to further biological contexts.

%\bibliography{refs}
%\bibliographystyle{iEEEtr}

\onecolumngrid
\appendix

\section{Direct Transfer (DT)}

Assume there are $N+1$ cells, one source cell and $N$ target cells. The source cell produces morphogen molecules at rate $\beta$. These molecules can then be sent to the $j$th target cell at rate $\gamma_{j}$, wherein they degrade with rate $\nu$. Let $m_{0}$ be the number of molecules in the source cell and $m_{j}$ be the number of molecules in the $j$th target cell. Approximating these values as continuous allows for the equations

\begin{equation}
\frac{\partial m_{0}}{\partial t} = \beta+\eta_{\beta}-\sum_{j}\left(\gamma_{j}m_{0}+\eta_{\gamma j}\right)
\label{dm0}
\end{equation}

\begin{equation}
\frac{\partial m_{j}}{\partial t} = \gamma_{j}m_{0}+\eta_{\gamma j}-\nu m_{j}-\eta_{\nu j},
\label{dmj}
\end{equation}

where the various $\eta$ terms are Gaussian white noise terms. For simplification, let $\Gamma = \sum_{j}\gamma_{j}$. With that, linearizing and separating Eqs. \ref{dm0} and \ref{dmj} into $m_{j} = \bar{m}_{j}+\delta m_{j}$ yields

\begin{subequations}
\begin{equation}
\frac{\partial \bar{m}_{0}}{\partial t} = \beta-\Gamma\bar{m}_{0}
\label{m0bar}
\end{equation}
\begin{equation}
\frac{\partial\delta m_{0}}{\partial t} = \eta_{\beta}-\sum_{j}\left(\gamma_{j}\delta m_{0}+\eta_{\gamma j}\right)
\label{deltam0}
\end{equation}
\label{linm0}
\end{subequations}

\begin{subequations}
\begin{equation}
\frac{\partial \bar{m}_{j}}{\partial t} = \gamma_{j}\bar{m}_{0}-\nu\bar{m}_{j}
\label{mjbar}
\end{equation}
\begin{equation}
\frac{\partial\delta m_{j}}{\partial t} = \gamma_{j}\delta m_{0}+\eta_{\gamma j}-\nu\delta m_{j}-\eta_{\nu j}.
\label{deltamj}
\end{equation}
\label{linmj}
\end{subequations}

Setting Eqs. \ref{m0bar} and \ref{mjbar} to 0 yields mean values for $m_{0}$ and $m_{j}$ to be

\begin{equation}
\bar{m}_{0} = \frac{\beta}{\Gamma}
\label{m0mean}
\end{equation}

\begin{equation}
\bar{m}_{j} = \frac{\gamma_{j}\bar{m}_{0}}{\nu} = \frac{\beta\gamma_{j}}{\Gamma\nu}.
\label{mjmean}
\end{equation}

From here, Eqs. \ref{deltam0} and \ref{deltamj} can be Fourier transformed to yield

\begin{equation}
-i\omega\tilde{\delta m_{0}} = \tilde{\eta}_{\beta}-\sum_{j}\left(\gamma_{j}\tilde{\delta m_{0}}+\tilde{\eta}_{\gamma j}\right) \implies \tilde{\delta m_{0}} = \frac{\tilde{\eta}_{\beta}-\sum_{j}\tilde{\eta}_{\gamma j}}{\Gamma-i\omega}
\label{m0FT}
\end{equation}

\begin{align}
&-i\omega\tilde{\delta m_{j}} = \gamma_{j}\tilde{\delta m_{0}}+\tilde{\eta}_{\gamma j}-\nu\tilde{\delta m_{j}}-\tilde{\eta}_{\nu j} \nonumber\\
&\implies \tilde{\delta m_{j}} = \frac{\gamma_{j}\tilde{\delta m_{0}}+\tilde{\eta}_{\gamma j}-\tilde{\eta}_{\nu j}}{\nu-i\omega} = \frac{1}{\nu-i\omega}\left(\frac{\gamma_{j}\tilde{\eta}_{\beta}}{\Gamma-i\omega}-\sum_{s}\left(\frac{\gamma_{j}}{\Gamma-i\omega}-\delta_{js}\right)\tilde{\eta}_{\gamma s}-\tilde{\eta}_{\nu j}\right).
\label{mjFT}
\end{align}

Since each $\eta$ term represents the noise from a simple linear reaction, they must be independent of each other, and their cross spectrums must be $\delta$-correlated in time and frequency space with magnitudes equal to the mean propensity of their respective reactions. This implies

\begin{equation}
\left\langle\tilde{\eta}_{\beta}^{*}\left(\omega'\right)\tilde{\eta}_{\beta}\left(\omega\right)\right\rangle = \beta\left(2\pi\delta\left(\omega-\omega'\right)\right)
\label{etabeta}
\end{equation}

\begin{equation}
\left\langle\tilde{\eta}_{\gamma k}^{*}\left(\omega'\right)\tilde{\eta}_{\gamma j}\left(\omega\right)\right\rangle = \gamma_{j}\bar{m}_{0}\delta_{jk}\left(2\pi\delta\left(\omega-\omega'\right)\right)
\label{etagammajk}
\end{equation}

\begin{equation}
\left\langle\tilde{\eta}_{\nu k}^{*}\left(\omega'\right)\tilde{\eta}_{\nu j}\left(\omega\right)\right\rangle = \nu\bar{m}_{j}\delta_{jk}\left(2\pi\delta\left(\omega-\omega'\right)\right) = \gamma_{j}\bar{m}_{0}\delta_{jk}\left(2\pi\delta\left(\omega-\omega'\right)\right).
\label{etanujk}
\end{equation}

These allow for the cross spectrum of $m_{0}$ to be

\begin{align}
\left\langle\tilde{\delta m_{0}}^{*}\left(\omega'\right)\tilde{\delta m_{0}}\left(\omega\right)\right\rangle &= \frac{\left\langle\tilde{\eta}_{\beta}^{*}\left(\omega'\right)\tilde{\eta}_{\beta}\left(\omega\right)\right\rangle+\sum_{j,k}\left\langle\tilde{\eta}_{\gamma k}^{*}\left(\omega'\right)\tilde{\eta}_{\gamma j}\left(\omega\right)\right\rangle}{\left(\Gamma-i\omega\right)\left(\Gamma+i\omega'\right)} \nonumber\\
&= \left(2\pi\delta\left(\omega-\omega'\right)\right)\frac{\beta+\bar{m}_{0}\sum_{j,k}\gamma_{j}\delta_{jk}}{\left(\Gamma-i\omega\right)\left(\Gamma+i\omega'\right)} \nonumber\\
&= \frac{2\beta}{\Gamma^{2}+\omega^{2}}\left(2\pi\delta\left(\omega-\omega'\right)\right).
\label{m0cross}
\end{align}

Utilizing the zero-frequency approximation here allows for the noise-to-signal ratio for $m_{0}$ to be

\begin{equation}
\frac{\delta m_{T0}^{2}}{\bar{m}_{0}^{2}} = \frac{2\beta}{\Gamma^{2}T}\left(\frac{\beta}{\Gamma}\right)^{-2} = \frac{2}{\bar{m}_{0}\Gamma T}.
\label{m0NSR}
\end{equation}

Moving on to $m_{j}$, Eqs. \ref{mjFT}-\ref{etanujk} allow for the cross spectrum between $m_{j}$ and $m_{k}$ to be

\begin{align}
&\left\langle\tilde{\delta m_{k}}^{*}\left(\omega'\right)\tilde{\delta m_{j}}\left(\omega\right)\right\rangle = \frac{1}{\left(\nu-i\omega\right)\left(\nu+i\omega'\right)}\left(\frac{\gamma_{j}\gamma_{k}\left\langle\tilde{\eta}_{\beta}^{*}\left(\omega'\right)\tilde{\eta}_{\beta}\left(\omega\right)\right\rangle}{\left(\Gamma-i\omega\right)\left(\Gamma+i\omega'\right)}+\left\langle\tilde{\eta}_{\nu k}^{*}\left(\omega'\right)\tilde{\eta}_{\nu j}\left(\omega\right)\right\rangle\right. \nonumber\\
&\quad\quad \left.+\sum_{s,t}\left(\frac{\gamma_{j}}{\Gamma-i\omega}-\delta_{js}\right)\left(\frac{\gamma_{k}}{\Gamma+i\omega'}-\delta_{kt}\right)\left\langle\tilde{\eta}_{\gamma t}^{*}\left(\omega'\right)\tilde{\eta}_{\gamma s}\left(\omega\right)\right\rangle \vphantom{\frac{\gamma_{j}\gamma_{k}\left\langle\tilde{\eta}_{\beta}^{*}\left(\omega'\right)\tilde{\eta}_{\beta}\left(\omega\right)\right\rangle}{\left(\Gamma-i\omega\right)\left(\Gamma+i\omega'\right)}}\right) \nonumber\\
&\quad = \frac{2\pi\delta\left(\omega-\omega'\right)}{\left(\nu-i\omega\right)\left(\nu+i\omega'\right)}\left(\frac{\beta\gamma_{j}\gamma_{k}}{\left(\Gamma-i\omega\right)\left(\Gamma+i\omega'\right)}+\gamma_{j}\bar{m}_{0}\delta_{jk} \vphantom{\sum_{s,t}\left(\frac{\gamma_{j}}{\Gamma-i\omega}-\delta_{js}\right)}\right. \nonumber\\
&\quad\quad \left.+\bar{m}_{0}\sum_{s,t}\left(\frac{\gamma_{j}}{\Gamma-i\omega}-\delta_{js}\right)\left(\frac{\gamma_{k}}{\Gamma+i\omega'}-\delta_{kt}\right)\gamma_{s}\delta_{st}\right) \nonumber\\
&\quad = \frac{2\pi\delta\left(\omega-\omega'\right)}{\left(\nu-i\omega\right)\left(\nu+i\omega'\right)}\left(\underbrace{\frac{\beta\gamma_{j}\gamma_{k}}{\left(\Gamma-i\omega\right)\left(\Gamma+i\omega'\right)}}_{\text{production}}+\underbrace{\gamma_{j}\bar{m}_{0}\delta_{jk}}_{\text{degradation}}\right. \nonumber\\
&\quad\quad \left.+\bar{m}_{0}\sum_{s}\left(\underbrace{\frac{\gamma_{j}\gamma_{k}}{\left(\Gamma-i\omega\right)\left(\Gamma+i\omega'\right)}}_{\text{production correlation}}-\underbrace{\frac{\gamma_{j}\delta_{ks}}{\Gamma-i\omega}-\frac{\gamma_{k}\delta_{js}}{\Gamma+i\omega'}}_{\text{transport correlation}}+\underbrace{\delta_{js}\delta_{ks}}_{\text{transport}}\right)\gamma_{s}\right) \nonumber\\
&\quad = \frac{2\pi\delta\left(\omega-\omega'\right)}{\left(\nu-i\omega\right)\left(\nu+i\omega'\right)}\left(\frac{2\beta\gamma_{j}\gamma_{k}}{\left(\Gamma-i\omega\right)\left(\Gamma+i\omega'\right)}+2\gamma_{j}\bar{m}_{0}\delta_{jk}-\gamma_{j}\gamma_{k}\bar{m}_{0}\left(\frac{1}{\Gamma-i\omega}+\frac{1}{\Gamma+i\omega'}\right)\right) \nonumber\\
&\quad = \frac{2\pi\delta\left(\omega-\omega'\right)}{\left(\nu-i\omega\right)\left(\nu+i\omega'\right)}\left(\frac{2\beta\gamma_{j}\gamma_{k}}{\left(\Gamma-i\omega\right)\left(\Gamma+i\omega'\right)}+2\gamma_{j}\bar{m}_{0}\delta_{jk}-\gamma_{j}\gamma_{k}\bar{m}_{0}\frac{2\Gamma-i\left(\omega-\omega'\right)}{\left(\Gamma-i\omega\right)\left(\Gamma+i\omega'\right)}\right) \nonumber\\
&\quad = 2\gamma_{j}\bar{m}_{0}\delta_{jk}\frac{2\pi\delta\left(\omega-\omega'\right)}{\left(\nu-i\omega\right)\left(\nu+i\omega'\right)}.
\label{mjkcross}
\end{align}

Once again utilizing the zero-frequency approximation allows for the noise-to-signal ratio for $m_{j}$ to be

\begin{equation}
\frac{\delta m_{Tj}^{2}}{\bar{m}_{j}^{2}} = \frac{2\gamma_{j}\bar{m}_{0}}{\nu^{2}\bar{m}_{j}^{2}T} = \frac{2}{\bar{m}_{j}\nu T}.
\label{mjNSRA}
\end{equation}

Eq. \ref{mjNSRA} gives the full form the noise-to-signal ratio for the $j$th target cell. Interestingly, it contains no cross correlation terms with the source cell or any other target cell nor any term inherited from the production noise in the source cell. The reason for this can be seen midway through the derivation of Eq. \ref{mjkcross}. The noise terms that explicitly arise from the production, degredation, and transport processes are highlighted as well as the positive cross correlation term that arises from production noise and negative cross correlation term that arises from transport noise. The negative transport correlation is seen to cancel both the production noise and production correlation, thus leaving the cell with only a degredation noise term and an identical noise term from the transport process.

\section{Synthesis-Diffusion-Clearance (SDC)}

We can now compare this model to the SDC model. We will look at diffusion in a multitude of different spaces with different dimensions as well as morphogen sources that span a multitude of different dimensions. In each case, the sources will secrete morphogen molecules into a density field $\rho$ which must follow

\begin{equation}
\frac{\partial\rho}{\partial t} = D\nabla^{2}\rho+\eta_{D}-\nu\rho-\eta_{\nu}+\left(\beta+\eta_{\beta}\right)\delta^{SP-SO}\left(\vec{x}\right),
\label{rhoeq}
\end{equation}

where $SP$ is the number of spatial dimensions, $SO$ is the dimensionality of the source, and $\nabla^{2}$ is taken over all $SP$ dimensions. Of important note is that $\delta^{SP-SO}\left(\vec{x}\right)$ is a $\delta$ function only in the {\bf last} $SP-SO$ dimensions of the space. So, for example, if there was a 1 dimensional source in 3 dimensional space, then $\delta^{3-1}\left(\vec{x}\right)$ would be a $\delta$ function in the $\hat{y}$ and $\hat{z}$ directions but not the $\hat{x}$ direction. This means that $\beta$ and $\eta_{\beta}$ will have units of $T^{-1}L^{-SO}$, where $T$ is time and $L$ is space.

We can now assume $\rho$ has reached a steady state and separate it into $\rho = \bar{\rho}+\delta\rho$, which in turn allows Eq. \ref{rhoeq} to separate into

\begin{equation}
0 = D\nabla^{2}\bar{\rho}-\nu\bar{\rho}+\beta\delta^{SP-SO}\left(\vec{x}\right)
\label{rhobar}
\end{equation}

\begin{equation}
\frac{\partial\delta\rho}{\partial t} = D\nabla^{2}\delta\rho+\eta_{D}-\nu\delta\rho-\eta_{\nu}+\eta_{\beta}\delta^{SP-SO}\left(\vec{x}\right).
\label{deltarho}
\end{equation}

Fourier transforming Eq. \ref{rhobar} in space and dividing it by $\nu$ then yields

\begin{equation}
0 = -\lambda^{2}\abs{\vec{k}}^{2}\tilde{\bar{\rho}}-\tilde{\bar{\rho}}+\frac{\beta\lambda^{2}}{D}\left(2\pi\right)^{SO}\delta^{SO}\left(\vec{k}\right) \implies \tilde{\bar{\rho}} = \frac{\beta\lambda^{2}}{D}\frac{\left(2\pi\right)^{SO}\delta^{SO}\left(\vec{k}\right)}{1+\lambda^{2}\abs{\vec{k}}^{2}},
\label{rhobarFT}
\end{equation}

where

\begin{equation}
\lambda = \sqrt{\frac{D}{\nu}}.
\label{lambdadef}
\end{equation}

Of similarly important note is that $\delta^{SO}\left(\vec{k}\right)$ is a $\delta$ function only in the {\bf first} $SO$ dimensions of $k$-space. So in the 1 dimensional source, 3 dimensional space example $\delta^{SO}\left(\vec{k}\right)$ would be a $\delta$ function in the $\hat{x}$ direction of $k$-space but not the $\hat{y}$ or $\hat{z}$ directions.

This allows $\bar{\rho}$ to be written as

\begin{align}
\bar{\rho}\left(\vec{x}\right) &= \int\frac{d^{SP}k}{\left(2\pi\right)^{SP}}e^{-i\vec{k}\cdot\vec{x}}\tilde{\bar{\rho}}\left(\vec{k}\right) = \frac{\beta\lambda^{2}}{D}\int\frac{d^{SP}k}{\left(2\pi\right)^{SP}}e^{-i\vec{k}\cdot\vec{x}}\frac{\left(2\pi\right)^{SO}\delta^{SO}\left(\vec{k}\right)}{1+\lambda^{2}\abs{\vec{k}}^{2}} \nonumber\\
&= \frac{\beta\lambda^{2}}{D}\int\frac{d^{SP-SO}k}{\left(2\pi\right)^{SP-SO}}e^{-i\vec{k}\cdot\vec{x}}\frac{1}{1+\lambda^{2}\abs{\vec{k}}^{2}} = \frac{\beta\lambda^{2-\left(SP-SO\right)}}{D}P_{SP-SO}\left(\frac{\abs{\vec{x}}}{\lambda}\right),
\label{rhobarform}
\end{align}

where

\begin{equation}
P_{N}\left(x\right) = \int\frac{d^{N}u}{\left(2\pi\right)^{N}}e^{-i\vec{u}\cdot\vec{x}}\frac{1}{1+\abs{\vec{u}}^{2}}.
\label{Pdef}
\end{equation}

It is important to note that $P_{N}$ does not integrate over all available dimensions, but only over the last $N$ dimensions of the space. This in turn means that its argument can only depend on the last $N$ dimensions of any input vector. Returning to the 1 dimensional source, 3 dimensional space example, $P_{3-1}\left(\abs{\vec{x}}/\lambda\right)$ should only take the $y$ and $z$ components of $\vec{x}$ into account. The $x$ component is made irrelevant by the translational symmetry of the system along the $x$-axis.

Moving on to the noise terms, Eq. \ref{deltarho} can be Fourier transformed in space and time to yield

\begin{equation}
-i\omega\tilde{\delta\rho} = -D\abs{\vec{k}}^{2}\tilde{\delta\rho}+\tilde{\eta}_{D}-\nu\tilde{\delta\rho}-\tilde{\eta}_{\nu}+\tilde{\eta}_{\beta} \implies \tilde{\delta\rho} = \frac{\tilde{\eta}_{D}-\tilde{\eta}_{\nu}+\tilde{\eta}_{\beta}}{\nu\left(1+\lambda^{2}\abs{\vec{k}}^{2}-i\frac{\omega}{\nu}\right)},
\label{deltarhoFT}
\end{equation}

where $\eta_{\beta}\left(\vec{k},\omega\right)$ depends only on the first $SO$ dimensions of $k$-space. Assuming the $\eta$ terms are all independent of each other allows the cross spectrum of $\rho$ to be

\begin{align}
&\left\langle\tilde{\delta\rho}^{*}\left(\vec{k}',\omega'\right)\tilde{\delta\rho}\left(\vec{k},\omega\right)\right\rangle = \frac{1}{\nu^{2}\left(1+\lambda^{2}\abs{\vec{k}}^{2}-i\frac{\omega}{\nu}\right)\left(1+\lambda^{2}\abs{\vec{k}'}^{2}+i\frac{\omega'}{\nu}\right)} \nonumber\\
&\quad \cdot\left(\left\langle\tilde{\eta}_{D}^{*}\left(\vec{k}',\omega'\right)\tilde{\eta}_{D}\left(\vec{k},\omega\right)\right\rangle+\left\langle\tilde{\eta}_{\nu}^{*}\left(\vec{k}',\omega'\right)\tilde{\eta}_{\nu}\left(\vec{k},\omega\right)\right\rangle+\left\langle\tilde{\eta}_{\beta}^{*}\left(\vec{k}',\omega'\right)\tilde{\eta}_{\beta}\left(\vec{k},\omega\right)\right\rangle\right).
\label{rhocross1}
\end{align}

The cross spectrum of $\eta_{D}$ can be obtained from its correlation function, which is known from \cite{gardiner2004Hosm} to take the form

\begin{equation}
\left\langle\eta_{D}\left(\vec{x}',t\right)\eta_{D}\left(\vec{x},t\right)\right\rangle = 2D\delta\left(t-t'\right)\vec{\nabla}\cdot\vec{\nabla}'\left(\bar{\rho}\left(\vec{x}\right)\delta^{SP}\left(\vec{x}-\vec{x}'\right)\right).
\label{etaDcorr}
\end{equation}

Fourier transforming Eq. \ref{etaDcorr} can be easily performed due to the $\delta$ functions, integrating the spatial terms by parts, and utilizing Eq. \ref{rhobarFT} to yield

\begin{align}
&\left\langle\tilde{\eta}_{D}^{*}\left(\vec{k}',\omega'\right)\tilde{\eta}_{D}\left(\vec{k},\omega\right)\right\rangle = \int d^{SP}xd^{SP}x'dtdt'e^{i\vec{k}\cdot\vec{x}}e^{-i\vec{k}'\cdot\vec{x}'}e^{i\omega t}e^{-i\omega't'}\left\langle\eta_{D}\left(\vec{x}',t\right)\eta_{D}\left(\vec{x},t\right)\right\rangle \nonumber\\
&\quad = 2D\int d^{SP}xd^{SP}x'dtdt'e^{i\vec{k}\cdot\vec{x}}e^{-i\vec{k}'\cdot\vec{x}'}e^{i\omega t}e^{-i\omega't'}\delta\left(t-t'\right)\vec{\nabla}\cdot\vec{\nabla}'\left(\bar{\rho}\left(\vec{x}\right)\delta^{SP}\left(\vec{x}-\vec{x}'\right)\right) \nonumber\\
&\quad = 2D\left(2\pi\delta\left(\omega-\omega'\right)\right)\int d^{SP}xd^{SP}x'e^{i\vec{k}\cdot\vec{x}}e^{-i\vec{k}'\cdot\vec{x}'}\vec{\nabla}\cdot\vec{\nabla}'\left(\bar{\rho}\left(\vec{x}\right)\delta^{SP}\left(\vec{x}-\vec{x}'\right)\right) \nonumber\\
&\quad = 2D\left(2\pi\delta\left(\omega-\omega'\right)\right)\int d^{SP}xd^{SP}x'\bar{\rho}\left(\vec{x}\right)\delta^{SP}\left(\vec{x}-\vec{x}'\right)\vec{\nabla}\cdot\vec{\nabla}'\left(e^{i\vec{k}\cdot\vec{x}}e^{-i\vec{k}'\cdot\vec{x}'}\right) \nonumber\\
&\quad = 2D\vec{k}\cdot\vec{k}'\left(2\pi\delta\left(\omega-\omega'\right)\right)\int d^{SP}xd^{SP}x'\bar{\rho}\left(\vec{x}\right)\delta^{SP}\left(\vec{x}-\vec{x}'\right)e^{i\vec{k}\cdot\vec{x}}e^{-i\vec{k}'\cdot\vec{x}'} \nonumber\\
&\quad = 2D\vec{k}\cdot\vec{k}'\left(2\pi\delta\left(\omega-\omega'\right)\right)\int d^{SP}x\bar{\rho}\left(\vec{x}\right)e^{i\vec{x}\left(\vec{k}-\vec{k}'\right)} \nonumber\\
&\quad = 2D\vec{k}\cdot\vec{k}'\tilde{\bar{\rho}}\left(\vec{k}-\vec{k}'\right)\left(2\pi\delta\left(\omega-\omega'\right)\right) \nonumber\\
&\quad = \frac{2\lambda^{2}\vec{k}\cdot\vec{k}'}{1+\lambda^{2}\abs{\vec{k}-\vec{k}'}^{2}}\left(\beta\left(2\pi\right)^{SO+1}\delta\left(\omega-\omega'\right)\delta^{SO}\left(\vec{k}-\vec{k}'\right)\right).
\label{etaDcross}
\end{align}

Moving on to $\eta_{\nu}$, its correlation function is known to take to form

\begin{equation}
\left\langle\eta_{\nu}\left(\vec{x}',t\right)\eta_{\nu}\left(\vec{x},t\right)\right\rangle = \nu\bar{\rho}\left(\vec{x}\right)\delta\left(t-t'\right)\delta^{SP}\left(\vec{x}-\vec{x}'\right).
\label{etanucorr}
\end{equation}

Fourier transforming Eq. \ref{etanucorr} is again easily performed due to the $\delta$ functions and Eq. \ref{rhobarFT}. This yields

\begin{align}
&\left\langle\tilde{\eta}_{\nu}^{*}\left(\vec{k}',\omega'\right)\tilde{\eta}_{\nu}\left(\vec{k},\omega\right)\right\rangle = \int d^{SP}xd^{SP}x'dtdt'e^{i\vec{k}\cdot\vec{x}}e^{-i\vec{k}'\cdot\vec{x}'}e^{i\omega t}e^{-i\omega't'}\left\langle\eta_{\nu}\left(\vec{x}',t\right)\eta_{\nu}\left(\vec{x},t\right)\right\rangle \nonumber\\
&\quad = \nu\int d^{SP}xd^{SP}x'dtdt'e^{i\vec{k}\cdot\vec{x}}e^{-i\vec{k}'\cdot\vec{x}'}e^{i\omega t}e^{-i\omega't'}\bar{\rho}\left(\vec{x}\right)\delta\left(t-t'\right)\delta^{SP}\left(\vec{x}-\vec{x}'\right) \nonumber\\
&\quad = \nu\left(2\pi\delta\left(\omega-\omega'\right)\right)\int d^{SP}xd^{SP}x'e^{i\vec{k}\cdot\vec{x}}e^{-i\vec{k}'\cdot\vec{x}'}\bar{\rho}\left(\vec{x}\right)\delta^{SP}\left(\vec{x}-\vec{x}'\right) \nonumber\\
&\quad = \nu\left(2\pi\delta\left(\omega-\omega'\right)\right)\int d^{SP}xe^{i\vec{x}\cdot\left(\vec{k}-\vec{k}'\right)}\bar{\rho}\left(\vec{x}\right) \nonumber\\
&\quad = \nu\tilde{\bar{\rho}}\left(\vec{k}-\vec{k}'\right)\left(2\pi\delta\left(\omega-\omega'\right)\right) \nonumber\\
&\quad = \frac{1}{1+\lambda^{2}\abs{\vec{k}-\vec{k}'}^{2}}\left(\beta\left(2\pi\right)^{SO+1}\delta\left(\omega-\omega'\right)\delta^{SO}\left(\vec{k}-\vec{k}'\right)\right).
\label{etanucross}
\end{align}

Finally, the cross spectrum of $\eta_{\beta}$ must be $\delta$ correlated in $\omega$-space as well as all source dimensions of $k$-space since it is merely a uniform production term that does not depend of $\bar{\rho}$. This yields

\begin{equation}
\left\langle\tilde{\eta}_{\beta}^{*}\left(\vec{k}',\omega'\right)\tilde{\eta}_{\beta}\left(\vec{k},\omega\right)\right\rangle = \beta\left(2\pi\right)^{SO+1}\delta\left(\omega-\omega'\right)\delta^{SO}\left(\vec{k}-\vec{k}'\right).
\label{etabetacross}
\end{equation}

Combining Eqs. \ref{rhocross1}, \ref{etaDcross}, \ref{etanucross}, and \ref{etabetacross} then yields

\begin{align}
&\left\langle\tilde{\delta\rho}^{*}\left(\vec{k}',\omega'\right)\tilde{\delta\rho}\left(\vec{k},\omega\right)\right\rangle = \frac{\beta\left(2\pi\right)^{SO+1}\delta\left(\omega-\omega'\right)\delta^{SO}\left(\vec{k}-\vec{k}'\right)}{\nu^{2}\left(1+\lambda^{2}\abs{\vec{k}}^{2}-i\frac{\omega}{\nu}\right)\left(1+\lambda^{2}\abs{\vec{k}'}^{2}+i\frac{\omega'}{\nu}\right)} \nonumber\\
&\quad\quad \cdot\left(\frac{2\lambda^{2}\vec{k}\cdot\vec{k}'}{1+\lambda^{2}\abs{\vec{k}-\vec{k}'}^{2}}+\frac{1}{1+\lambda^{2}\abs{\vec{k}-\vec{k}'}^{2}}+1\right) \nonumber\\
&\quad = \frac{\beta\left(2\pi\right)^{SO+1}\delta\left(\omega-\omega'\right)\delta^{SO}\left(\vec{k}-\vec{k}'\right)}{\nu^{2}\left(1+\lambda^{2}\abs{\vec{k}}^{2}-i\frac{\omega}{\nu}\right)\left(1+\lambda^{2}\abs{\vec{k}'}^{2}+i\frac{\omega'}{\nu}\right)}\frac{2+\lambda^{2}\left(\abs{\vec{k}}^{2}+\abs{\vec{k}'}^{2}\right)}{1+\lambda^{2}\abs{\vec{k}-\vec{k}'}^{2}} \nonumber\\
&\quad = \frac{\tilde{\bar{\rho}}\left(\vec{k}-\vec{k}'\right)\left(2\pi\delta\left(\omega-\omega'\right)\right)\left(2+\lambda^{2}\left(\abs{\vec{k}}^{2}+\abs{\vec{k}'}^{2}\right)\right)}{\nu\left(1+\lambda^{2}\abs{\vec{k}}^{2}-i\frac{\omega}{\nu}\right)\left(1+\lambda^{2}\abs{\vec{k}'}^{2}+i\frac{\omega'}{\nu}\right)}.
\label{rhocross2}
\end{align}

We now define $m$ as

\begin{equation}
m\left(\vec{x},t\right) = \int_{V\left(a\right)}d^{SP}r\rho\left(\vec{x}+\vec{r},t\right),
\label{mdef}
\end{equation}

where $V\left(a\right)$ is a $SP$-dimensional sphere with radius $a$. This allows the mean value of $m$ to be written as

\begin{align}
\bar{m}\left(\vec{x}\right) &= \int_{V\left(a\right)}d^{SP}r\bar{\rho}\left(\vec{x}+\vec{r}\right) = \frac{\beta\lambda^{2-\left(SP-SO\right)}}{D}\int_{V\left(a\right)}d^{SP}rP_{SP-SO}\left(\frac{\abs{\vec{x}+\vec{r}}}{\lambda}\right) \nonumber\\
&= \frac{\beta\lambda^{SO}}{\nu}M_{SP-SO,SP}\left(\frac{\abs{\vec{x}}}{\lambda},\frac{a}{\lambda}\right),
\label{mbar}
\end{align}

where

\begin{equation}
M_{N,N'}\left(x,y\right) = \int_{V\left(y\right)}d^{N'}uP_{N}\left(\abs{\vec{x}+\vec{u}}\right).
\label{Mfuncdef}
\end{equation}

Since $P_{N}\left(\abs{\vec{x}}\right)$ can only depend on the last $N$ dimensions of its input vectors, the same must be true of $M_{N,N'}$. From here we define $S\left(\vec{x}\right)$ as the 0-frequency limit of the cross spectrum in $\omega$-space of $m$. This allows it to take the form 

\begin{align}
S\left(\vec{x}\right) &= \lim_{\omega\to 0}\int\frac{d\omega'}{2\pi}\left\langle\tilde{\delta m}^{*}\left(\vec{x},\omega'\right)\tilde{\delta m}\left(\vec{x},\omega\right)\right\rangle \nonumber\\
&= \lim_{\omega\to 0}\int\frac{d\omega'}{2\pi}\int_{V\left(a\right)}d^{SP}rd^{SP}r'\int\frac{d^{SP}k}{\left(2\pi\right)^{SP}}\frac{d^{SP}k'}{\left(2\pi\right)^{SP}}e^{-i\vec{k}\cdot\left(\vec{x}+\vec{r}\right)}e^{i\vec{k}'\cdot\left(\vec{x}+\vec{r}'\right)}\left\langle\tilde{\delta\rho}^{*}\left(\vec{k}',\omega'\right)\tilde{\delta\rho}\left(\vec{k},\omega\right)\right\rangle \nonumber\\
&= \frac{1}{\left(2\pi\right)^{2SP}\nu}\int_{V\left(a\right)}d^{SP}rd^{SP}r'\int d^{SP}kd^{SP}k'e^{-i\vec{k}\cdot\left(\vec{x}+\vec{r}\right)}e^{i\vec{k}'\cdot\left(\vec{x}+\vec{r}'\right)} \nonumber\\
&\quad\quad \cdot\frac{\tilde{\bar{\rho}}\left(\vec{k}-\vec{k}'\right)\left(2+\lambda^{2}\left(\abs{\vec{k}}^{2}+\abs{\vec{k}'}^{2}\right)\right)}{\left(1+\lambda^{2}\abs{\vec{k}}^{2}\right)\left(1+\lambda^{2}\abs{\vec{k}'}^{2}\right)} \nonumber\\
&= \frac{1}{\left(2\pi\right)^{2SP}\nu}\int_{V\left(a\right)}d^{SP}rd^{SP}r'\int d^{SP}kd^{SP}k'd^{SP}ze^{-i\vec{k}\cdot\left(\vec{x}+\vec{r}\right)}e^{i\vec{k}'\cdot\left(\vec{x}+\vec{r}'\right)}e^{i\vec{z}\cdot\left(\vec{k}-\vec{k}'\right)} \nonumber\\
&\quad\quad \cdot\bar{\rho}\left(\vec{z}\right)\frac{2+\lambda^{2}\left(\abs{\vec{k}}^{2}+\abs{\vec{k}'}^{2}\right)}{\left(1+\lambda^{2}\abs{\vec{k}}^{2}\right)\left(1+\lambda^{2}\abs{\vec{k}'}^{2}\right)} \nonumber\\
&= \frac{1}{\left(2\pi\right)^{2SP}\nu}\int_{V\left(a\right)}d^{SP}rd^{SP}r'\int d^{SP}kd^{SP}k'd^{SP}ze^{-i\vec{k}\cdot\left(\vec{x}+\vec{r}-\vec{z}\right)}e^{i\vec{k}'\cdot\left(\vec{x}+\vec{r}'-\vec{z}\right)} \nonumber\\
&\quad\quad \cdot\bar{\rho}\left(\vec{z}\right)\left(\frac{1}{1+\lambda^{2}\abs{\vec{k}}^{2}}+\frac{1}{1+\lambda^{2}\abs{\vec{k}'}^{2}}\right) \nonumber\\
&= \frac{1}{\left(2\pi\right)^{2SP}\nu}\int_{V\left(a\right)}d^{SP}rd^{SP}r'\int d^{SP}z\bar{\rho}\left(\vec{z}\right)\left(\int d^{SP}ke^{-i\vec{k}\cdot\left(\vec{x}+\vec{r}-\vec{z}\right)}\frac{\left(2\pi\right)^{SP}\delta^{SP}\left(\vec{x}+\vec{r}'-\vec{z}\right)}{1+\lambda^{2}\abs{\vec{k}}^{2}}\right. \nonumber\\
&\quad\quad \left.+\int d^{SP}k'e^{i\vec{k}'\cdot\left(\vec{x}+\vec{r}'-\vec{z}\right)}\frac{\left(2\pi\right)^{SP}\delta^{SP}\left(\vec{x}+\vec{r}-\vec{z}\right)}{1+\lambda^{2}\abs{\vec{k}}^{2}}\right) \nonumber\\
&= \frac{\beta\lambda^{2-\left(SP-SO\right)}}{D\nu\lambda^{SP}}\int_{V\left(a\right)}d^{SP}rd^{SP}r'\int d^{SP}zP_{SP-SO}\left(\frac{\abs{\vec{z}}}{\lambda}\right) \nonumber\\
&\quad\quad \cdot\left(\delta^{SP}\left(\vec{x}+\vec{r}'-\vec{z}\right)P_{SP}\left(\frac{\abs{\vec{x}+\vec{r}-\vec{z}}}{\lambda}\right)+\delta^{SP}\left(\vec{x}+\vec{r}-\vec{z}\right)P_{SP}\left(\frac{\abs{\vec{x}+\vec{r}'-\vec{z}}}{\lambda}\right)\right) \nonumber\\
&= \frac{\beta\lambda^{4-\left(2SP-SO\right)}}{D^{2}}\int_{V\left(a\right)}d^{SP}rd^{SP}r'P_{SP}\left(\frac{\abs{\vec{r}-\vec{r}'}}{\lambda}\right)\left(P_{SP-SO}\left(\frac{\abs{\vec{x}+\vec{r}}}{\lambda}\right)+P_{SP-SO}\left(\frac{\abs{\vec{x}+\vec{r}'}}{\lambda}\right)\right) \nonumber\\
&= \frac{\beta\lambda^{4-\left(SP-SO\right)}}{D^{2}}\left(\int_{V\left(a\right)}d^{SP}rM_{SP,SP}\left(\frac{\abs{\vec{r}}}{\lambda},\frac{a}{\lambda}\right)P_{SP-SO}\left(\frac{\abs{\vec{x}+\vec{r}}}{\lambda}\right)\right. \nonumber\\
&\quad +\left.\int_{V\left(a\right)}d^{SP}r'M_{SP,SP}\left(\frac{\abs{\vec{r}'}}{\lambda},\frac{a}{\lambda}\right)P_{SP-SO}\left(\frac{\abs{\vec{x}+\vec{r}'}}{\lambda}\right)\right) \nonumber\\
&= \frac{2\beta\lambda^{4-\left(SP-SO\right)}}{D^{2}}\int_{V\left(a\right)}d^{SP}rM_{SP,SP}\left(\frac{\abs{\vec{r}}}{\lambda},\frac{a}{\lambda}\right)P_{SP-SO}\left(\frac{\abs{\vec{x}+\vec{r}}}{\lambda}\right) \nonumber\\
&= \frac{2\beta\lambda^{SO}}{\nu^{2}}\Sigma_{SP-SO,SP}\left(\frac{\abs{\vec{x}}}{\lambda},\frac{a}{\lambda}\right),
\label{mcross}
\end{align}

where

\begin{equation}
\Sigma_{N,N'}\left(x,y\right) = \int_{V\left(y\right)}d^{N'}uM_{N',N'}\left(u,y\right)P_{N}\left(\abs{\vec{x}+\vec{u}}\right).
\label{Sigmadef}
\end{equation}

Wherein once again only the last $N$ dimensions of the input vectors can be taken into account. Combining Eqs. \ref{mbar} and \ref{mcross} yields the full 0-frequency noise-to-signal ratio of $m$ to be

\begin{equation}
\frac{\delta m_{T}^{2}}{\bar{m}^{2}} = \frac{S}{\bar{m}^{2}T} = \frac{2}{\lambda^{SO}\beta T}\frac{\Sigma_{SP-SO,SP}\left(\frac{\abs{\vec{x}}}{\lambda},\frac{a}{\lambda}\right)}{\left(M_{SP-SO,SP}\left(\frac{\abs{\vec{x}}}{\lambda},\frac{a}{\lambda}\right)\right)^{2}}.
\label{mNSR}
\end{equation}

With Eq. \ref{mNSR}, once the forms of $P_{N}$, $M_{N,N'}$, and $\Sigma_{N,N'}$ are determined for a given $SP$ and $SO$, the full form of the noise-to-signal ratio can be found.

\section{Comparing DT and SDC}

The DT and SDC models can be directly compared to each other to determine which one has a lesser noise-to-signal ratio. To do so, we assume there are $N$ targets cells in each system. Let the positions in space of the target cells in the diffusion model be denoted by $\vec{x}_{j}$ for $j\in\lbrack 1,N\rbrack$. From here, each $\gamma_{j}$ in the DT model is scaled such that $\bar{m}\left(\vec{x}_{j}\right)$ from the SDC model and $\bar{m}_{j}$ from the DT model are equivalent. In particular, this means

\begin{equation}
\frac{\beta_{d}\lambda^{SO}}{\nu_{d}}M_{SP-SO,SP}\left(\frac{\abs{\vec{x}_{j}}}{\lambda},\frac{a}{\lambda}\right) = \frac{\beta_{c}\gamma_{j}}{\nu_{c}\Gamma},
\label{meqm}
\end{equation}

where the $d$ subscript denoted that parameter belongs to the SDC model and the $c$ subscript denotes that parameter belongs to the DT model. Normalizing both sides of Eq. \ref{meqm} by summing over $j$ then yields

\begin{equation}
\frac{M_{SP-SO,SP}\left(\frac{\abs{\vec{x}_{j}}}{\lambda},\frac{a}{\lambda}\right)}{\sum_{j=1}^{N}M_{SP-SO,SP}\left(\frac{\abs{\vec{x}_{j}}}{\lambda},\frac{a}{\lambda}\right)} = \frac{\gamma_{j}}{\Gamma}.
\label{gammaratio}
\end{equation}

Let $R_{SP-SO,SP}\left(\{\vec{x}_{j}\},N\right)$ be the ratio of the noise-to-signal ratios of the SDC model to the DT model. Combining this with Eqs. \ref{mjNSRA}, \ref{mNSR}, and \ref{gammaratio} then yields

\begin{align}
R_{SP-SO,SP}\left(\{\vec{x}_{j}\},N\right) &= \frac{\beta_{c}T}{\lambda^{SO}\beta_{d}T}\frac{\Sigma_{SP-SO,SP}\left(\frac{\abs{\vec{x}_{j}}}{\lambda},\frac{a}{\lambda}\right)}{\left(M_{SP-SO,SP}\left(\frac{\abs{\vec{x}_{j}}}{\lambda},\frac{a}{\lambda}\right)\right)^{2}}\frac{\gamma_{j}}{\Gamma} \nonumber\\
&= \frac{\beta_{c}}{\lambda^{SO}\beta_{d}}\frac{\Sigma_{SP-SO,SP}\left(\frac{\abs{\vec{x}_{j}}}{\lambda},\frac{a}{\lambda}\right)}{M_{SP-SO,SP}\left(\frac{\abs{\vec{x}_{j}}}{\lambda},\frac{a}{\lambda}\right)\sum_{k=1}^{N}M_{SP-SO,SP}\left(\frac{\abs{\vec{x}_{k}}}{\lambda},\frac{a}{\lambda}\right)}.
\label{Rdef}
\end{align}

When $R<1$ the SDC model has less noise, while when $R>1$ the DT model has less noise.

\section{1D space, 0D source}

To begin, we start with the simple scenario in which $SP=1$ and $SO=0$. This allows $P_{1}$, $M_{1,1}$, and $\Sigma_{1,1}$ to take the forms

\begin{equation}
P_{1}\left(x\right) = \int\frac{du}{2\pi}e^{-iux}\frac{1}{1+u^{2}} = \frac{1}{2}e^{-\abs{x}}
\label{P1}
\end{equation}

\begin{align}
M_{1,1}\left(x,y\right) &= \int_{-y}^{y}duP_{1}\left(\abs{x+u}\right) = \frac{1}{2}\int_{-y}^{y}due^{-\abs{x+u}} \nonumber\\
&= \begin{cases}
1-e^{-y}\cosh\left(x\right) & x<y \\
e^{-x}\sinh\left(y\right) & x\ge y \end{cases}
\label{M11}
\end{align}

\begin{align}
\Sigma_{1,1}\left(x,y\right) &= \int_{-y}^{y}duM_{1,1}\left(u,y\right)P_{1}\left(\abs{x+u}\right) = \frac{1}{2}\int_{-y}^{y}du\left(1-e^{-y}\cosh\left(u\right)\right)e^{-\abs{x+u}} \nonumber\\
&= \begin{cases}
1-\frac{1}{4}e^{-y}\left(\left(5+2y-e^{-2y}\right)\cosh\left(x\right)-2x\sinh\left(x\right)\right) & x<y \\
\frac{1}{4}e^{-x}\left(4\sinh\left(y\right)-e^{-y}\left(2y+\sinh\left(2y\right)\right)\right) & x\ge y \end{cases}.
\label{Sigma11}
\end{align}

Now assume that there are $N$ target cells placed in a line such that $x_{j}=2aj$. This allows $R_{1,1}\left(\{x_{j}\},N\right)$ to take the form

\begin{align}
R_{1,1}\left(\{x_{j}\},N\right) &= \frac{\beta_{c}}{\beta_{d}}\frac{\frac{1}{4}e^{-\frac{2aj}{\lambda}}\left(4\sinh\left(\frac{a}{\lambda}\right)-e^{-\frac{a}{\lambda}}\left(2\frac{a}{\lambda}+\sinh\left(2\frac{a}{\lambda}\right)\right)\right)}{e^{-\frac{2aj}{\lambda}}\sinh\left(\frac{a}{\lambda}\right)\sum_{k=1}^{N}e^{-\frac{2ak}{\lambda}}\sinh\left(\frac{a}{\lambda}\right)} \nonumber\\
&= \frac{\beta_{c}}{\beta_{d}}\frac{4\sinh\left(\frac{a}{\lambda}\right)-e^{-\frac{a}{\lambda}}\left(2\frac{a}{\lambda}+\sinh\left(2\frac{a}{\lambda}\right)\right)}{4e^{-\frac{a}{\lambda}\left(N+1\right)}\sinh\left(\frac{a}{\lambda}\right)\sinh\left(\frac{aN}{\lambda}\right)}
\label{R11}
\end{align}

If the production rates are scaled such that $\beta_{c}=\beta_{d}$ then there exists a critical value of $\hat{\lambda}=\lambda/a$ at $\hat{\lambda}^{c}\approx 2.346900061$. For any $\hat{\lambda}<\hat{\lambda}^{c}$, $R_{1,1}$ must be greater than 1, which in turn implies the DT model must produce lesser noise regardless of population size. The $R_{1,1}=1$ curve in $(N,\hat{\lambda})$ space with $\beta_{c}=\beta_{d}$ is plotted in Fig. 2A (1D case) of the main text.

Eqs. \ref{M11} and \ref{Sigma11} can also be put into Eq. \ref{mNSR} to obtain the noise-to-signal ratio of the SDC model. However, these expressions do not make clear the distinction between the components that arise from the three distinct noise terms seen in Eq. \ref{rhocross1}. We now rederive Eq. \ref{mNSR} specifically for the $SP=1$ and $SO=0$ geometry as an example of how these terms can be held separate. To do so, we go back to Eq. \ref{rhocross2} and insert it into the second line of Eq. \ref{mcross} along with Eqs. \ref{etaDcross}, \ref{etanucross}, and \ref{etabetacross} to obtain

\begin{align}
S\left(x\right) &= \lim_{\omega\to 0}\int\frac{d\omega'}{2\pi}\int_{V\left(a\right)}drdr'\int\frac{dk}{2\pi}\frac{dk'}{2\pi}e^{-ik\left(x+r\right)}e^{ik'\left(x+r'\right)} \nonumber\\
&\quad\quad \cdot\frac{\left\langle\tilde{\eta}_{D}^{*}\left(k',\omega'\right)\tilde{\eta}_{D}\left(k,\omega\right)\right\rangle+\left\langle\tilde{\eta}_{\nu}^{*}\left(k',\omega'\right)\tilde{\eta}_{\nu}\left(k,\omega\right)\right\rangle+\left\langle\tilde{\eta}_{\beta}^{*}\left(\omega'\right)\tilde{\eta}_{\beta}\left(\omega\right)\right\rangle}{\nu^{2}\left(1+\lambda^{2}k^{2}-i\frac{\omega}{\nu}\right)\left(1+\lambda^{2}{k'}^{2}+i\frac{\omega'}{\nu}\right)} \nonumber\\
&= \lim_{\omega\to 0}\int\frac{d\omega'}{2\pi}\int_{-a}^{a}drdr'\int\frac{dk}{2\pi}\frac{dk'}{2\pi}e^{-ik\left(x+r\right)}e^{ik'\left(x+r'\right)} \nonumber\\
&\quad\quad\cdot\frac{2\pi\beta\delta\left(\omega-\omega'\right)}{\nu^{2}\left(1+\lambda^{2}k^{2}-i\frac{\omega}{\nu}\right)\left(1+\lambda^{2}{k'}^{2}+i\frac{\omega'}{\nu}\right)}\left(\frac{2\lambda^{2}kk'}{1+\lambda^{2}\left(k-k'\right)^{2}}+\frac{1}{1+\lambda^{2}\left(k-k'\right)^{2}}+1\right) \nonumber\\
&= \frac{\beta}{\nu^{2}}\left(I_{D}\left(x\right)+I_{\nu}\left(x\right)+I_{\beta}\left(x\right)\right),
\label{S111}
\end{align}

where

\begin{equation}
I_{D}\left(x\right) = \frac{1}{\left(2\pi\right)^{2}}\int_{-a}^{a}drdr'\int dkdk'\frac{e^{-ik\left(x+r\right)}e^{ik'\left(x+r'\right)}}{\left(1+\lambda^{2}k^{2}\right)\left(1+\lambda^{2}{k'}^{2}\right)}\frac{2\lambda^{2}kk'}{1+\lambda^{2}\left(k-k'\right)^{2}}
\label{IDdef}
\end{equation}

\begin{equation}
I_{\nu}\left(x\right) = \frac{1}{\left(2\pi\right)^{2}}\int_{-a}^{a}drdr'\int dkdk'\frac{e^{-ik\left(x+r\right)}e^{ik'\left(x+r'\right)}}{\left(1+\lambda^{2}k^{2}\right)\left(1+\lambda^{2}{k'}^{2}\right)}\frac{1}{1+\lambda^{2}\left(k-k'\right)^{2}}
\label{Inudef}
\end{equation}

\begin{equation}
I_{\beta}\left(x\right) = \frac{1}{\left(2\pi\right)^{2}}\int_{-a}^{a}drdr'\int dkdk'\frac{e^{-ik\left(x+r\right)}e^{ik'\left(x+r'\right)}}{\left(1+\lambda^{2}k^{2}\right)\left(1+\lambda^{2}{k'}^{2}\right)}.
\label{Ibetadef}
\end{equation}

Utilizing Eq. \ref{rhobarFT} allows for $I_{D}\left(x\right)$ to be solved, yielding

\begin{align}
I_{D}\left(x\right) &= \frac{1}{\left(2\pi\right)^{2}}\int_{-a}^{a}drdr'\int dkdk'\frac{e^{-ik\left(x+r\right)}e^{ik'\left(x+r'\right)}}{\left(1+\lambda^{2}k^{2}\right)\left(1+\lambda^{2}{k'}^{2}\right)}\left(2\lambda^{2}kk'\frac{\nu\tilde{\bar{\rho}}\left(k-k'\right)}{\beta}\right) \nonumber\\
&= \frac{2\lambda^{2}\nu}{\left(2\pi\right)^{2}\beta}\int_{-a}^{a}drdr'\int dkdk'dz\frac{kk'\bar{\rho}\left(z\right)}{\left(1+\lambda^{2}k^{2}\right)\left(1+\lambda^{2}{k'}^{2}\right)}e^{-ik\left(x+r\right)}e^{ik'\left(x+r'\right)}e^{iz\left(k-k'\right)} \nonumber\\
&= \frac{2\lambda^{2}\nu}{\left(2\pi\right)^{2}\beta}\int_{-a}^{a}drdr'\frac{\partial}{\partial r}\frac{\partial}{\partial r'}\int dkdk'dz\frac{\bar{\rho}\left(z\right)}{\left(1+\lambda^{2}k^{2}\right)\left(1+\lambda^{2}{k'}^{2}\right)}e^{-ik\left(x+r-z\right)}e^{ik'\left(x+r'-z\right)} \nonumber\\
&= \frac{2\lambda^{2}\nu}{\left(2\pi\right)^{2}\beta}\int_{-a}^{a}drdr'\frac{\partial}{\partial r}\frac{\partial}{\partial r'}\int dkdk'dz\bar{\rho}\left(z\right)\left(\frac{\nu\tilde{\bar{\rho}}\left(k\right)}{\beta}\right)\left(\frac{\nu\tilde{\bar{\rho}}\left(k'\right)}{\beta}\right)e^{-ik\left(x+r-z\right)}e^{ik'\left(x+r'-z\right)} \nonumber\\
&= \frac{2\lambda^{2}\nu^{3}}{\beta^{3}}\int_{-a}^{a}drdr'\frac{\partial}{\partial r}\frac{\partial}{\partial r'}\int dz\bar{\rho}\left(z\right)\bar{\rho}\left(x+r-z\right)\bar{\rho}\left(x+r'-z\right) \nonumber\\
&= \frac{2\lambda^{2}\nu^{3}}{\beta^{3}}\int dz\bar{\rho}\left(z\right)\left(\bar{\rho}\left(x+a-z\right)-\bar{\rho}\left(x-a-z\right)\right)^{2}.
\label{IDsol1}
\end{align}

Eqs. \ref{rhobarform} and \ref{P1} can then be used to obtain

\begin{align}
I_{D}\left(x\right) &= \frac{1}{4\lambda}\int dze^{-\frac{\abs{z}}{\lambda}}\left(e^{-\frac{\abs{x+a-z}}{\lambda}}-e^{-\frac{\abs{x-a-z}}{\lambda}}\right)^{2} \nonumber\\
&= \begin{cases}
\frac{2}{3}e^{-\frac{a}{\lambda}}\cosh\left(\frac{\abs{x}}{\lambda}\right)\left(1+e^{-2\frac{a}{\lambda}}\right)-e^{-2\frac{a}{\lambda}}\left(1+\frac{1}{3}\cosh\left(2\frac{\abs{x}}{\lambda}\right)\right) & \abs{x}<a \\
\frac{2}{3}e^{-\frac{\abs{x}}{\lambda}}\sinh\left(\frac{a}{\lambda}\right)\left(1-e^{-2\frac{a}{\lambda}}\right)-\frac{1}{3}e^{-2\frac{\abs{x}}{\lambda}}\left(\cosh\left(2\frac{a}{\lambda}\right)-1\right) & \abs{x}\ge a \end{cases}.
\label{IDsol2}
\end{align}

The same technique can be used to solve for $I_{\nu}\left(x\right)$ to yeild

\begin{align}
I_{\nu}\left(x\right) &= \frac{1}{\left(2\pi\right)^{2}}\int_{-a}^{a}drdr'\int dkdk'\frac{e^{-ik\left(x+r\right)}e^{ik'\left(x+r'\right)}}{\left(1+\lambda^{2}k^{2}\right)\left(1+\lambda^{2}{k'}^{2}\right)}\left(\frac{\nu\tilde{\bar{\rho}}\left(k-k'\right)}{\beta}\right) \nonumber\\
&= \frac{\nu}{\left(2\pi\right)^{2}\beta}\int_{-a}^{a}drdr'\int dkdk'dz\frac{\bar{\rho}\left(z\right)}{\left(1+\lambda^{2}k^{2}\right)\left(1+\lambda^{2}{k'}^{2}\right)}e^{-ik\left(x+r\right)}e^{ik'\left(x+r'\right)}e^{iz\left(k-k'\right)} \nonumber\\
&= \frac{\nu}{\left(2\pi\right)^{2}\beta}\int_{-a}^{a}drdr'\int dkdk'dz\bar{\rho}\left(z\right)\left(\frac{\nu\tilde{\bar{\rho}}\left(k\right)}{\beta}\right)\left(\frac{\nu\tilde{\bar{\rho}}\left(k'\right)}{\beta}\right)e^{-ik\left(x+r-z\right)}e^{ik'\left(x+r'-z\right)} \nonumber\\
&= \frac{\nu^{3}}{\beta^{3}}\int_{-a}^{a}drdr'\int dz\bar{\rho}\left(z\right)\bar{\rho}\left(x+r-z\right)\bar{\rho}\left(x+r'-z\right) \nonumber\\
&= \frac{1}{8\lambda^{3}}\int_{-a}^{a}drdr'\int dze^{-\frac{\abs{z}}{\lambda}}e^{-\frac{\abs{x+r-z}}{\lambda}}e^{-\frac{\abs{x+r'-z}}{\lambda}} \nonumber\\
&= \begin{cases}
1+\frac{\abs{x}}{\lambda}e^{-\frac{a}{\lambda}}\sinh\left(\frac{\abs{x}}{\lambda}\right)-\frac{1}{6}e^{-2\frac{a}{\lambda}}\left(\cosh\left(2\frac{\abs{x}}{\lambda}\right)-3\right)-\left(\frac{7}{6}+\frac{a}{\lambda}+\frac{1}{6}e^{-2\frac{a}{\lambda}}\right)e^{-\frac{a}{\lambda}}\cosh\left(\frac{\abs{x}}{\lambda}\right) & \abs{x}<a \\
\frac{2}{3}e^{-\frac{\abs{x}}{\lambda}}\sinh\left(\frac{a}{\lambda}\right)+e^{-\frac{\abs{x}}{\lambda}}e^{-\frac{a}{\lambda}}\left(\frac{1}{6}\sinh\left(2\frac{a}{\lambda}\right)-\frac{a}{\lambda}\right)-\frac{1}{6}e^{-2\frac{\abs{x}}{\lambda}}\left(\cosh\left(2\frac{a}{\lambda}\right)-1\right) & \abs{x}\ge a \end{cases}.
\label{Inusol}
\end{align}

Lastly, $I_{\beta}\left(x\right)$ can be easily solved to yield

\begin{align}
I_{\beta}\left(x\right) &= \frac{1}{\left(2\pi\right)^{2}}\int_{-a}^{a}drdr'\int dkdk'\frac{e^{-ik\left(x+r\right)}e^{ik'\left(x+r'\right)}}{\left(1+\lambda^{2}k^{2}\right)\left(1+\lambda^{2}{k'}^{2}\right)} \nonumber\\
&= \frac{1}{\left(2\pi\right)^{2}}\int_{-a}^{a}drdr'\int dkdk'\left(\frac{\nu\tilde{\bar{\rho}}\left(k\right)}{\beta}\right)\left(\frac{\nu\tilde{\bar{\rho}}\left(k'\right)}{\beta}\right)e^{-ik\left(x+r\right)}e^{ik'\left(x+r'\right)} \nonumber\\
&= \frac{\nu^{2}}{\beta^{2}}\int_{-a}^{a}drdr'\bar{\rho}\left(x+r\right)\bar{\rho}\left(x+r'\right) \nonumber\\
&= \frac{1}{4\lambda^{2}}\int_{-a}^{a}drdr'e^{-\frac{\abs{x+r}}{\lambda}}e^{-\frac{\abs{x+r'}}{\lambda}} \nonumber\\
&= \begin{cases}
\left(1-e^{-\frac{a}{\lambda}}\cosh\left(\frac{\abs{x}}{\lambda}\right)\right)^{2} & \abs{x}<a \\
\left(e^{-\frac{\abs{x}}{\lambda}}\sinh\left(\frac{a}{\lambda}\right)\right)^{2} & \abs{x}\ge a \end{cases}.
\label{Ibetasol}
\end{align}

Combining Eqs. \ref{mbar}, \ref{mNSR}, \ref{S111}, and \ref{IDsol2}-\ref{Ibetasol} gives the full form of the $SP=1$, $SO=0$ noise-to-signal ratio.

\begin{equation}
\frac{\delta m_{T}^{2}}{\bar{m}^{2}} = \frac{1}{\lambda^{SO}\beta T}\frac{I_{D}\left(x\right)+I_{\nu}\left(x\right)+I_{\beta}\left(x\right)}{\left(M_{1,1}\left(\frac{\abs{x}}{\lambda},\frac{a}{\lambda}\right)\right)^{2}}
\label{mNSR11}
\end{equation}

Each $I\left(x\right)$ term represents the noise inherited from a different source. Importantly, each different $I\left(x\right)$ term is nonnegative for any value of $x$. This shows that unlike Eq. \ref{mjNSRA} in the DT model, there are no negative correlation terms to cancel any of the inherited noise terms.

\section{2D space, 0D source}

For $SP=2$ and $SO=0$, $P_{2}$, $M_{2,2}$, and $\Sigma_{2,2}$ each take the form

\begin{equation}
P_{2}\left(x\right) = \int\frac{d^{2}u}{\left(2\pi\right)^{2}}e^{-i\vec{u}\cdot\vec{x}}\frac{1}{1+\abs{\vec{u}}^{2}} = \frac{1}{2\pi}K_{0}\left(x\right)
\label{P2}
\end{equation}

\begin{align}
M_{2,2}\left(x,y\right) &= \int_{V\left(y\right)}d^{2}uP_{2}\left(\abs{\vec{x}+\vec{u}}\right) = \int_{V\left(y\right)}d^{2}u\int\frac{d^{2}u'}{\left(2\pi\right)^{2}}e^{-i\vec{u}'\cdot\left(\vec{x}+\vec{u}\right)}\frac{1}{1+\abs{\vec{u}'}^{2}} \nonumber\\
&= y\int_{0}^{\infty}du'\frac{J_{0}\left(xu'\right)J_{1}\left(yu'\right)}{1+{u'}^{2}}
\label{M22}
\end{align}

\begin{align}
\Sigma_{2,2}\left(x,y\right) &= \int_{V\left(y\right)}d^{2}uM_{2,2}\left(\abs{\vec{u}},y\right)P_{2}\left(\abs{\vec{x}+\vec{u}}\right) \nonumber\\
&= y\int_{V\left(y\right)}d^{2}u\int_{0}^{\infty}du'\int\frac{d^{2}u''}{\left(2\pi\right)^{2}}\frac{J_{0}\left(\abs{\vec{u}}u'\right)J_{1}\left(yu'\right)}{1+{u'}^{2}}\frac{e^{-i\vec{u}''\cdot\left(\vec{x}+\vec{u}\right)}}{1+\abs{\vec{u}''}^{2}} \nonumber\\
&= y^{2}\int_{0}^{\infty}du'du''\frac{u''J_{0}\left(xu''\right)J_{1}\left(yu'\right)\left(u'J_{0}\left(yu''\right)J_{1}\left(yu'\right)-u''J_{0}\left(yu'\right)J_{1}\left(yu''\right)\right)}{\left({u'}^{2}-{u''}^{2}\right)\left(1+{u'}^{2}\right)\left(1+{u''}^{2}\right)},
\label{Sigma22}
\end{align}

where $J_{n}\left(x\right)$ and $K_{n}\left(x\right)$ are the Bessel functions of the first kind and modified Bessel functions of the second kind respectively. Unfortunately, the complicated nature of Bessel functions makes the remaining integrals unsolvable analytically. Similar problems arise whenever $SP=2$ or $SP-SO=2$.

\section{3D space, 0D source}

For $SP=3$ and $SO=0$, $P_{3}$, $M_{3,3}$, and $\Sigma_{3,3}$ each take the form

\begin{equation}
P_{3}\left(x\right) = \int\frac{d^{3}u}{\left(2\pi\right)^{3}}e^{-i\vec{u}\cdot\vec{x}}\frac{1}{1+\abs{\vec{u}}^{2}} = \frac{1}{4\pi x}e^{-x}
\label{P3}
\end{equation}

\begin{align}
M_{3,3}\left(x,y\right) &= \int_{V\left(y\right)}d^{3}uP_{3}\left(\abs{\vec{x}+\vec{u}}\right) = \frac{1}{4\pi}\int_{V\left(y\right)}d^{3}u\frac{1}{\abs{\vec{x}+\vec{u}}}e^{-\abs{\vec{x}+\vec{u}}} \nonumber\\
&= \begin{cases}
1-\frac{1+y}{x}e^{-y}\sinh\left(x\right) & x<y \\
\frac{1}{x}e^{-x}\left(y\cosh\left(y\right)-\sinh\left(y\right)\right) & x\ge y \end{cases}
\label{M33}
\end{align}

\begin{align}
\Sigma_{3,3}\left(x,y\right) &= \int_{V\left(y\right)}d^{3}uM_{3,3}\left(\abs{\vec{u}},y\right)P_{3}\left(\abs{\vec{x}+\vec{u}}\right) \nonumber\\
&= \frac{1}{4\pi}\int_{V\left(y\right)}d^{3}u\left(1-\frac{1+y}{\abs{\vec{u}}}e^{-y}\sinh\left(\abs{\vec{u}}\right)\right)\frac{1}{\abs{\vec{x}+\vec{u}}}e^{-\abs{\vec{x}+\vec{u}}} \nonumber\\
&= \begin{cases}
1-\frac{1}{4x}e^{-y}\left(1+y\right)\left(\left(5+2y+e^{-2y}\right)\sinh\left(x\right)-2x\cosh\left(x\right)\right) & x<y \\
\frac{1}{4x}e^{-x}\left(4\left(y\cosh\left(y\right)-\sinh\left(y\right)\right)+e^{-y}\left(1+y\right)\left(2y-\sinh\left(2y\right)\right)\right) & x\ge y \end{cases}
\label{Sigma33}
\end{align}

This allows $R_{3,3}\left(\{\vec{x}_{j}\},N\right)$ with $\abs{\vec{x}_{j}} = 2aj$ to take the form

\begin{align}
R_{3,3}\left(\{\vec{x}_{j}\},N\right) &= \frac{\beta_{c}}{\beta_{d}}\frac{\frac{\lambda}{8aj}e^{-\frac{2aj}{\lambda}}\left(4\left(\frac{a}{\lambda}\cosh\left(\frac{a}{\lambda}\right)-\sinh\left(\frac{a}{\lambda}\right)\right)+e^{-\frac{a}{\lambda}}\left(1+\frac{a}{\lambda}\right)\left(2\frac{a}{\lambda}-\sinh\left(2\frac{a}{\lambda}\right)\right)\right)}{\frac{\lambda}{2aj}e^{-\frac{2aj}{\lambda}}\left(\frac{a}{\lambda}\cosh\left(\frac{a}{\lambda}\right)-\sinh\left(\frac{a}{\lambda}\right)\right)\sum_{k=1}^{N}\frac{\lambda}{2ak}e^{-\frac{2ak}{\lambda}}\left(\frac{a}{\lambda}\cosh\left(\frac{a}{\lambda}\right)-\sinh\left(\frac{a}{\lambda}\right)\right)} \nonumber\\
&= \frac{\beta_{c}}{\beta_{d}}\frac{4\left(\frac{a}{\lambda}\cosh\left(\frac{a}{\lambda}\right)-\sinh\left(\frac{a}{\lambda}\right)\right)+e^{-\frac{a}{\lambda}}\left(1+\frac{a}{\lambda}\right)\left(2\frac{a}{\lambda}-\sinh\left(2\frac{a}{\lambda}\right)\right)}{4\left(\frac{a}{\lambda}\cosh\left(\frac{a}{\lambda}\right)-\sinh\left(\frac{a}{\lambda}\right)\right)^{2}\sum_{k=1}^{N}\frac{\lambda}{2ak}e^{-\frac{2ak}{\lambda}}}
\label{R33}
\end{align}

Again scaling the production rate such that $\beta_{c}=\beta_{d}$, one critical value of $\hat{\lambda}$ can be recovered at $\hat{\lambda}^{c}\approx 18.85244498$. For any $\hat{\lambda}<\hat{\lambda}^{c}$, $R_{3,3}$ must necessarily be greater than 1, which in turn implies the DT model must produce lesser noise. There is also a lower bound on $N$ at $N^{c}=5$. For any $N\le N^{c}$, $R_{3,3}$ must necessarily be greater than 1, which again implies the DT model must produce lesser noise.

\section{2D space, 1D source}

For $SP=2$, $SO=1$, $P_{1}$ and $M_{2,2}$ are known from Eqs. \ref{P1} and \ref{M22}. This leaves $M_{1,2}$ and $\Sigma_{1,2}$ to take the forms

\begin{equation}
M_{1,2}\left(x,y\right) = \int_{V\left(y\right)}d^{2}uP_{1}\left(\abs{\vec{x}+\vec{u}}\right) = \frac{1}{2}\int_{0}^{y}du\int_{0}^{2\pi}d\theta ue^{-\abs{x_{2}+u_{2}}} = e^{-\abs{x_{2}}}\int_{0}^{2\pi}d\theta\frac{1-e^{-y\sin\left(\theta\right)}\left(1+y\sin\left(\theta\right)\right)}{2\left(\sin\left(\theta\right)\right)^{2}}
\label{M12}
\end{equation}

\begin{equation}
\Sigma_{1,2}\left(x,y\right) = \int_{V\left(y\right)}d^{2}uM_{2,2}\left(\abs{\vec{u}},y\right)P_{1}\left(\abs{\vec{x}+\vec{u}}\right) = \frac{y}{2}\int_{0}^{y}du\int_{0}^{2\pi}d\theta\int_{0}^{\infty}du'u\frac{J_{0}\left(uu'\right)J_{1}\left(yu'\right)}{1+{u'}^{2}}e^{-\abs{x_{2}+u\sin\left(\theta\right)}}
\label{Sigma12}
\end{equation}

Unfortunately, the remaining integrals are unsolvable analytically. They can, however, be calculated numerically, and by assuming the source line in the SDC model is a line of cells, the production rate of each individual cell in the line can be made equal to that of the source cell in the DT model by setting $\beta_{c}=2a\beta_{d}$. The numerically calculated $R_{1,2}=1$ curve in $(N,\hat{\lambda})$ space with $\beta_{c}=2a\beta_{d}$ is plotted in Fig. 2A (2D case) of the main text.

\section{3D space, 2D source}

For $SP=3$, $SO=2$, $P_{1}$ and $M_{3,3}$ are known from Eqs. \ref{P1} and \ref{M33}. This leaves $M_{1,3}$ and $\Sigma_{1,3}$ to take the forms

\begin{align}
M_{1,3}\left(x,y\right) &= \int_{V\left(y\right)}d^{3}uP_{1}\left(\abs{\vec{x}+\vec{u}}\right) = \frac{1}{2}\int_{V\left(y\right)}d^{3}ue^{-\abs{x_{3}+u_{3}}} \nonumber\\
&= 2\pi\begin{cases}
e^{-y}\left(1+y\right)\cosh\left(x\right)+\frac{y^{2}-x^{2}}{2}-1 & x<y \\
e^{-x}\left(y\cosh\left(y\right)-\sinh\left(y\right)\right) & x\ge y \end{cases}
\label{M13}
\end{align}

\begin{align}
\Sigma_{1,3}\left(x,y\right) &= \int_{V\left(y\right)}d^{3}uM_{3,3}\left(\abs{\vec{u}},y\right)P_{1}\left(\abs{\vec{x}+\vec{u}}\right) = \frac{1}{2}\int_{V\left(y\right)}d^{3}u\left(1-\frac{1+y}{\abs{\vec{u}}}e^{-y}\sinh\left(\abs{\vec{u}}\right)\right)e^{-\abs{x_{3}+u_{3}}} \nonumber\\
&=2\pi\begin{cases}
e^{-y}\left(1+y\right)\left(\frac{7+2y+e^{-2y}}{4}\cosh\left(x\right)-\frac{x}{2}\sinh\left(x\right)-\cosh\left(y\right)\right)+\frac{y^{2}-x^{2}}{2}-1 & x<y \\
e^{-x}\left(\frac{4y^{2}+5y-1}{8}e^{-y}+\frac{1+y}{8}e^{-3y}+\frac{3y}{4}\cosh\left(y\right)-\frac{5}{4}\sinh\left(y\right)\right) & x\ge y \end{cases}
\label{Sigma13}
\end{align}

This allows $R_{1,3}\left(\{\vec{x}_{j}\},N\right)$ with $\abs{\vec{x}_{j}} = 2aj$ to take the form

\begin{align}
&R_{1,3}\left(\{\vec{x}_{j}\},N\right) \nonumber\\
&= \frac{\beta_{c}}{\lambda^{2}\beta_{d}}\frac{2\pi e^{-\frac{2aj}{\lambda}}\left(\frac{1}{8}\left(4\left(\frac{a}{\lambda}\right)^{2}+5\frac{a}{\lambda}-1\right)e^{-\frac{a}{\lambda}}+\frac{1}{8}\left(1+\frac{a}{\lambda}\right)e^{-3\frac{a}{\lambda}}+\frac{3a}{4\lambda}\cosh\left(\frac{a}{\lambda}\right)-\frac{5}{4}\sinh\left(\frac{a}{\lambda}\right)\right)}{\left(2\pi\right)^{2}e^{-\frac{2aj}{\lambda}}\left(\frac{a}{\lambda}\cosh\left(\frac{a}{\lambda}\right)-\sinh\left(\frac{a}{\lambda}\right)\right)\sum_{k=1}^{N}e^{-\frac{2ak}{\lambda}}\left(\frac{a}{\lambda}\cosh\left(\frac{a}{\lambda}\right)-\sinh\left(\frac{a}{\lambda}\right)\right)} \nonumber\\
&= \frac{\beta_{c}}{\lambda^{2}\beta_{d}}\frac{\sinh\left(\frac{a}{\lambda}\right)\left(\frac{1}{8}\left(4\left(\frac{a}{\lambda}\right)^{2}+5\frac{a}{\lambda}-1\right)e^{-\frac{a}{\lambda}}+\frac{1}{8}\left(1+\frac{a}{\lambda}\right)e^{-3\frac{a}{\lambda}}+\frac{3a}{4\lambda}\cosh\left(\frac{a}{\lambda}\right)-\frac{5}{4}\sinh\left(\frac{a}{\lambda}\right)\right)}{2\pi e^{-\frac{a}{\lambda}\left(N+1\right)}\sinh\left(\frac{aN}{\lambda}\right)\left(\frac{a}{\lambda}\cosh\left(\frac{a}{\lambda}\right)-\sinh\left(\frac{a}{\lambda}\right)\right)^{2}}
\label{R13}
\end{align}

By assuming the source plane in the SDC model is a triangular lattice of cells, the production rate of each individual cell in the lattice can be made equal to that of the source cell in the DT model by setting $\beta_{c}=2a^{2}\sqrt{3}\beta_{d}$. There then exists a critical value of $\hat{\lambda}=\lambda/a$ at $\hat{\lambda}^{c}\approx 1.279786379$. For any $\hat{\lambda}<\hat{\lambda}^{c}$, $R_{1,1}$ must be greater than 1, which in turn implies the DT model must produce lesser noise regardless of population size. The $R_{1,3}=1$ curve in $(N,\hat{\lambda})$ space with $\beta_{c}=2a^{2}\sqrt{3}\beta_{d}$ is plotted in Fig. 2A (3D case) of the main text.

\section{Comparison to experimental data}

To compare our theory to experimental data, we focus on ten of the morphogens presented in Table 1 of \cite{kicheva2012investigating} and obtain data from the references therein. For Bicoid, we obtain a value of $\lambda$ of $\sim$100$\mu$m from the text of \cite{gregor2007stability} with and error of $\pm10\mu$m from the finding in \cite{gregor2007probing} that cells have a $\sim$10\% error in measuring the Bicoid gradient. We then take the $a$ value of the {\it Drosophila} embryo cells that are subjected to the Bicoid gradient to be $\sim$2.8$\mu$m based on Fig. 3A of \cite{gregor2007probing}. This value of $a$ is also used for Dorsal as measurements of both Bicoid and Dorsal occur in the {\it Drosophila} embryo at nuclear cycle 14. For the value of $\lambda$ for Dorsal, we use Fig. 3D from \cite{liberman2009quantitative} to obtain a full width at 60\% max of 45$\pm$10$\mu$m. Since this represents the width of Gaussian fit on both sides of the source whereas our model uses an exponential profile, we assume the appropriate $\lambda$ value for such an exponential fit would be half this value, 22.5$\pm$5$\mu$m.

For Dpp and Wg, \cite{kicheva2007kinetics} provides explicit measurements of $\lambda$ for each. These values are 20.2$\pm$5.7$\mu$m and 5.8$\pm$2.04$\mu$m respectively. For Hh, we use Fig. S2C in the supplementary material of \cite{wartlick2011dynamics} to determine $\lambda$ to be 8$\pm$3$\mu$m. Dpp, Wg, and Hh all occur in the wing disc during the third instar of the {\it Drosophila} development. As such, we use a common value of $a$ for all three. This value is taken to be 1.3$\mu$m based on the area of the cells being reported as 5.5$\pm$0.8$\mu$m$^{2}$ in the supplementary material of \cite{kicheva2007kinetics} and the assumption that the cells are circular.

The $\lambda$ value of Fgf8 is reported as being 197$\pm$7$\mu$m in \cite{yu2009fgf8}. Additionally, based off the scale bars seen in Fig. 2C-E of \cite{yu2009fgf8}, we estimate the value of $a$ for the cells to be $\sim$10$\mu$m. For the morphogens involved in the Nodal/Lefty system (cyclops, squint, lefty1, and lefty2), measurements of $\lambda$ for each are taken from Fig. 2C-F of \cite{muller2012differential} by observing where the average of the three curves crosses the 37\% of max threshold with error bars given by the width of the region in which the vertical error bars of each plot intersect this threshold line. We assume the $a$ value of each morphogen in the Nodal/Lefty system to be equivalent to the $a$ value of cells in the Fgf8 measurements performed in \cite{yu2009fgf8}. This is because the measurements made in \cite{muller2012differential} we taken during the blastula stage of the zebrafish development while measurements taken in \cite{yu2009fgf8} we taken in the sphere germ ring stage. These stages occur at $\sim$2.25 and $\sim$5.67 hpf respectively, but the blastula stage can last until $\sim$6 hpf based on the timeline of zebrafish development presented in \cite{kimmel1995stages}. As such, since there is potential overlap in the time frame of these two stages, we assume the cells maintain a relatively fixed size and thus that the value of $a$ for the Nodal/Lefty system can be taken as the same value of $a$ used for Fgf8.

The $\hat{\lambda}=\lambda/a$ value of these ten morphogens can be seen plotted in Fig. 2B of the main text.


\begin{thebibliography}{10}

\bibitem{dubuis2013positional}
J.~O. Dubuis, G.~Tka{\v{c}}ik, E.~F. Wieschaus, T.~Gregor, and W.~Bialek,
  ``Positional information, in bits,'' {\em Proceedings of the National Academy
  of Sciences}, vol.~110, no.~41, pp.~16301--16308, 2013.

\bibitem{erdmann2009role}
T.~Erdmann, M.~Howard, and P.~R. Ten~Wolde, ``Role of spatial averaging in the
  precision of gene expression patterns,'' {\em Physical review letters},
  vol.~103, no.~25, p.~258101, 2009.

\bibitem{gregor2007probing}
T.~Gregor, D.~W. Tank, E.~F. Wieschaus, and W.~Bialek, ``Probing the limits to
  positional information,'' {\em Cell}, vol.~130, no.~1, pp.~153--164, 2007.

\bibitem{houchmandzadeh2002establishment}
B.~Houchmandzadeh, E.~Wieschaus, and S.~Leibler, ``Establishment of
  developmental precision and proportions in the early drosophila embryo,''
  {\em Nature}, vol.~415, no.~6873, p.~798, 2002.

\bibitem{de2010precision}
A.~M. De~Lachapelle and S.~Bergmann, ``Precision and scaling in morphogen
  gradient read-out,'' {\em Molecular systems biology}, vol.~6, no.~1, p.~351,
  2010.

\bibitem{akiyama2015morphogen}
T.~Akiyama and M.~C. Gibson, ``Morphogen transport: theoretical and
  experimental controversies,'' {\em Wiley Interdisciplinary Reviews:
  Developmental Biology}, vol.~4, no.~2, pp.~99--112, 2015.

\bibitem{gierer1972theory}
A.~Gierer and H.~Meinhardt, ``A theory of biological pattern formation,'' {\em
  Kybernetik}, vol.~12, no.~1, pp.~30--39, 1972.

\bibitem{lander2002morphogen}
A.~D. Lander, Q.~Nie, and F.~Y. Wan, ``Do morphogen gradients arise by
  diffusion?,'' {\em Developmental cell}, vol.~2, no.~6, pp.~785--796, 2002.

\bibitem{muller2013morphogen}
P.~M{\"u}ller, K.~W. Rogers, R.~Y. Shuizi, M.~Brand, and A.~F. Schier,
  ``Morphogen transport,'' {\em Development}, vol.~140, no.~8, pp.~1621--1638,
  2013.

\bibitem{rogers2011morphogen}
K.~W. Rogers and A.~F. Schier, ``Morphogen gradients: from generation to
  interpretation,'' {\em Annual review of cell and developmental biology},
  vol.~27, pp.~377--407, 2011.

\bibitem{wilcockson2017control}
S.~G. Wilcockson, C.~Sutcliffe, and H.~L. Ashe, ``Control of signaling molecule
  range during developmental patterning,'' {\em Cellular and Molecular Life
  Sciences}, vol.~74, no.~11, pp.~1937--1956, 2017.

\bibitem{bressloff2018bidirectional}
P.~C. Bressloff and H.~Kim, ``Bidirectional transport model of morphogen
  gradient formation via cytonemes,'' {\em Physical biology}, 2018.

\bibitem{kornberg2014cytonemes}
T.~B. Kornberg and S.~Roy, ``Cytonemes as specialized signaling filopodia,''
  {\em Development}, vol.~141, no.~4, pp.~729--736, 2014.

\bibitem{driever1988bicoid}
W.~Driever and C.~N{\"u}sslein-Volhard, ``The bicoid protein determines
  position in the drosophila embryo in a concentration-dependent manner,'' {\em
  Cell}, vol.~54, no.~1, pp.~95--104, 1988.

\bibitem{berezhkovskii2011formation}
A.~M. Berezhkovskii, C.~Sample, and S.~Y. Shvartsman, ``Formation of morphogen
  gradients: Local accumulation time,'' {\em Physical Review E}, vol.~83,
  no.~5, p.~051906, 2011.

\bibitem{yu2009fgf8}
S.~R. Yu, M.~Burkhardt, M.~Nowak, J.~Ries, Z.~Petr{\'a}{\v{s}}ek, S.~Scholpp,
  P.~Schwille, and M.~Brand, ``Fgf8 morphogen gradient forms by a source-sink
  mechanism with freely diffusing molecules,'' {\em Nature}, vol.~461,
  no.~7263, p.~533, 2009.

\bibitem{bischoff2013cytonemes}
M.~Bischoff, A.-C. Gradilla, I.~Seijo, G.~Andr{\'e}s, C.~Rodr{\'\i}guez-Navas,
  L.~Gonz{\'a}lez-M{\'e}ndez, and I.~Guerrero, ``Cytonemes are required for the
  establishment of a normal hedgehog morphogen gradient in drosophila
  epithelia,'' {\em Nature cell biology}, vol.~15, no.~11, p.~1269, 2013.

\bibitem{huang2015myoblast}
H.~Huang and T.~B. Kornberg, ``Myoblast cytonemes mediate wg signaling from the
  wing imaginal disc and delta-notch signaling to the air sac primordium,''
  {\em Elife}, vol.~4, p.~e06114, 2015.

\bibitem{stanganello2016role}
E.~Stanganello and S.~Scholpp, ``Role of cytonemes in wnt transport,'' {\em J
  Cell Sci}, vol.~129, no.~4, pp.~665--672, 2016.

\bibitem{shvartsman2012mathematical}
S.~Y. Shvartsman and R.~E. Baker, ``Mathematical models of morphogen gradients
  and their effects on gene expression,'' {\em Wiley Interdisciplinary Reviews:
  Developmental Biology}, vol.~1, no.~5, pp.~715--730, 2012.

\bibitem{teimouri2015new}
H.~Teimouri and A.~B. Kolomeisky, ``New model for understanding mechanisms of
  biological signaling: Direct transport via cytonemes,'' {\em The journal of
  physical chemistry letters}, vol.~7, no.~1, pp.~180--185, 2015.

\bibitem{teimouri2016mechanisms}
H.~Teimouri and A.~B. Kolomeisky, ``Mechanisms of the formation of biological
  signaling profiles,'' {\em Journal of Physics A: Mathematical and
  Theoretical}, vol.~49, no.~48, p.~483001, 2016.

\bibitem{grimm2010modelling}
O.~Grimm, M.~Coppey, and E.~Wieschaus, ``Modelling the bicoid gradient,'' {\em
  Development}, vol.~137, no.~14, pp.~2253--2264, 2010.

\bibitem{gregor2007stability}
T.~Gregor, E.~F. Wieschaus, A.~P. McGregor, W.~Bialek, and D.~W. Tank,
  ``Stability and nuclear dynamics of the bicoid morphogen gradient,'' {\em
  Cell}, vol.~130, no.~1, pp.~141--152, 2007.

\bibitem{kicheva2007kinetics}
A.~Kicheva, P.~Pantazis, T.~Bollenbach, Y.~Kalaidzidis, T.~Bittig,
  F.~J{\"u}licher, and M.~Gonzalez-Gaitan, ``Kinetics of morphogen gradient
  formation,'' {\em Science}, vol.~315, no.~5811, pp.~521--525, 2007.

\bibitem{kanodia2009dynamics}
J.~S. Kanodia, R.~Rikhy, Y.~Kim, V.~K. Lund, R.~DeLotto,
  J.~Lippincott-Schwartz, and S.~Y. Shvartsman, ``Dynamics of the dorsal
  morphogen gradient,'' {\em Proceedings of the National Academy of Sciences},
  vol.~106, no.~51, pp.~21707--21712, 2009.

\bibitem{gillespie2000chemical}
D.~T. Gillespie, ``The chemical langevin equation,'' {\em The Journal of
  Chemical Physics}, vol.~113, no.~1, pp.~297--306, 2000.

\bibitem{berg1977physics}
H.~C. Berg and E.~M. Purcell, ``Physics of chemoreception,'' {\em Biophysical
  journal}, vol.~20, no.~2, pp.~193--219, 1977.

\bibitem{supp}
See Appendices.

\bibitem{gardiner2004Hosm}
C.~W. Gardiner, {\em Handbook of stochastic methods for physics, chemistry, and
  the natural sciences}.
\newblock Springer series in synergetics (Unnumbered), Berlin ; New York:
  Springer, 3rd ed..~ed., 2004.

\bibitem{fancher2017fundamental}
S.~Fancher and A.~Mugler, ``Fundamental limits to collective concentration
  sensing in cell populations,'' {\em Physical review letters}, vol.~118,
  no.~7, p.~078101, 2017.

\bibitem{varennes2017emergent}
J.~Varennes, S.~Fancher, B.~Han, and A.~Mugler, ``Emergent versus
  individual-based multicellular chemotaxis,'' {\em Physical review letters},
  vol.~119, no.~18, p.~188101, 2017.

\bibitem{kicheva2012investigating}
A.~Kicheva, T.~Bollenbach, O.~Wartlick, F.~J{\"u}licher, and
  M.~Gonzalez-Gaitan, ``Investigating the principles of morphogen gradient
  formation: from tissues to cells,'' {\em Current opinion in genetics \&
  development}, vol.~22, no.~6, pp.~527--532, 2012.

\bibitem{wartlick2011dynamics}
O.~Wartlick, P.~Mumcu, A.~Kicheva, T.~Bittig, C.~Seum, F.~J{\"u}licher, and
  M.~Gonzalez-Gaitan, ``Dynamics of dpp signaling and proliferation control,''
  {\em Science}, vol.~331, no.~6021, pp.~1154--1159, 2011.

\bibitem{liberman2009quantitative}
L.~M. Liberman, G.~T. Reeves, and A.~Stathopoulos, ``Quantitative imaging of
  the dorsal nuclear gradient reveals limitations to threshold-dependent
  patterning in drosophila,'' {\em Proceedings of the National Academy of
  Sciences}, vol.~106, no.~52, pp.~22317--22322, 2009.

\bibitem{muller2012differential}
P.~M{\"u}ller, K.~W. Rogers, B.~M. Jordan, J.~S. Lee, D.~Robson, S.~Ramanathan,
  and A.~F. Schier, ``Differential diffusivity of nodal and lefty underlies a
  reaction-diffusion patterning system,'' {\em Science}, vol.~336, no.~6082,
  pp.~721--724, 2012.

\bibitem{rogers2018nodal}
K.~W. Rogers and P.~M{\"u}ller, ``Nodal and bmp dispersal during early
  zebrafish development,'' {\em Developmental biology}, 2018.

\end{thebibliography}
\end{document}